# Particle Energization in Space Plasmas:
# Towards a Multi-Point, Multi-Scale Plasma Observatory

Spokesperson: Alessandro Retinò

Laboratoire de Physique des Plasmas – Centre National de la Recherche Scientifique
École Polytechnique, route de Saclay, 91128 Palaiseau, France
E-mail: alessandro.retino@lpp.polytechnique.fr
Phone: +33-1-69335929

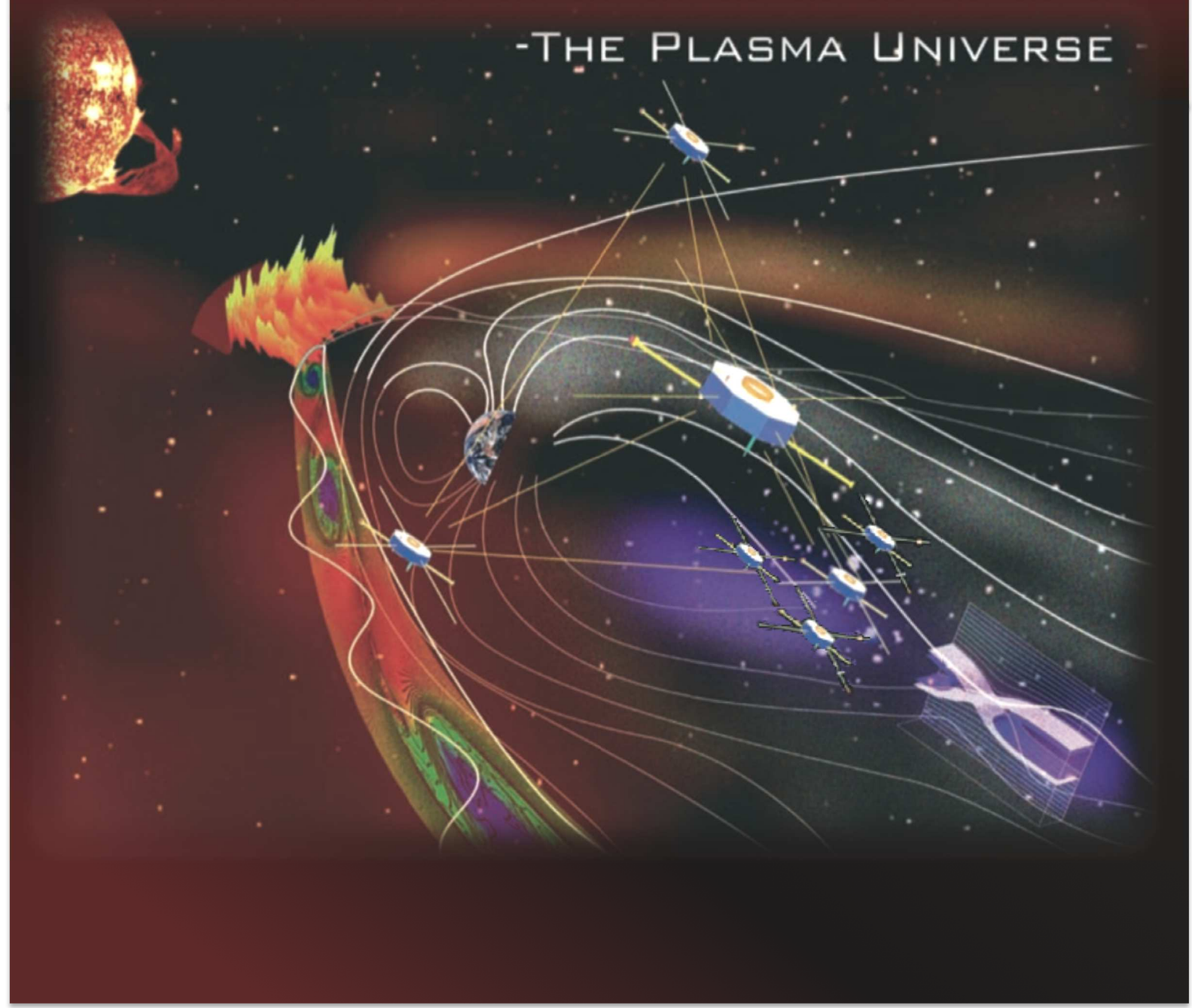

# Members of the Core Proposing Team

|    | Member                   | Affiliation                            |
|----|--------------------------|----------------------------------------|
| 1  | Alessandro Retinò        | LPP - CNRS, France                     |
| 2  | Yuri Khotyaintsev        | IRF, Uppsala, Sweden                   |
| 3  | Olivier Le Contel        | LPP - CNRS, France                     |
| 4  | Maria Federica Marcucci  | IAPS-INAF, Italy                       |
| 5  | Ferdinand Plaschke       | IWF-OEAW, Austria                      |
| 6  | Andris Vaivads           | KTH, Stockholm, Sweden                 |
| 7  | Vassilis Angelopoulos    | UCLA, USA                              |
| 8  | Pasquale Blasi           | Gran Sasso Science Institute, Italy    |
| 9  | Jim Burch                | SWRI, USA                              |
| 10 | Johan De Keyser          | BIRA-IASB, Belgium                     |
| 11 | Malcolm Dunlop           | RAL, UK                                |
| 12 | Lei Dai                  | NSSC, CAS, China                       |
| 13 | Jonathan Eastwood        | Imperial College London, UK            |
| 14 | Huishan Fu               | Beihang University, China              |
| 15 | Stein Haaland            | University of Bergen, Norway           |
| 16 | Masahiro Hoshino         | University of Tokyo, Japan             |
| 17 | Andreas Johlander        | University of Helsinki, Finland        |
| 18 | Larry Kepko              | GFSC, USA                              |
| 19 | Harald Kucharek          | University of New Hampshire, USA       |
| 20 | Gianni Lapenta           | KU Leuven, Belgium                     |
| 21 | Benoit Lavraud           | IRAP - CNRS, France                    |
| 22 | Olga Malandraki          | NOA, Greece                            |
| 23 | William Matthaeus        | University of Delaware, USA            |
| 24 | Kathryn McWilliams       | University of Saskatchewan, Canada     |
| 25 | Anatoli Petrukovich      | IKI, Russia                            |
| 26 | Jean-Louis Pinçon        | LPC2E - CNRS, France                   |
| 27 | Yoshifumi Saito          | ISAS-JAXA, Japan                       |
| 28 | Luca Sorriso-Valvo       | University of Calabria, Rende, Italy   |
| 29 | Rami Vainio              | University of Turku, Finland           |
| 30 | Bob Wimmer-Schweingruber | University of Kiel, Germany            |

Cover page: courtesy of Kentaro Tanaka and Maria Federica Marcucci.

# 1 Executive summary

**This White Paper outlines the importance of addressing the fundamental science theme "*How are charged particles energized in space plasmas*" through a future ESA mission. The White Paper presents five compelling science questions related to particle energization by shocks, reconnection, waves and turbulence, jets and their combinations. Answering these questions requires resolving scale coupling, nonlinearity and nonstationarity, which cannot be done with existing multi-point observations. In situ measurements from a multi-point, multi-scale L-class plasma observatory consisting of at least 7 spacecraft covering fluid, ion and electron scales are needed. The plasma observatory will enable a paradigm shift in our comprehension of particle energization and space plasma physics in general, with very important impact on solar and astrophysical plasmas. It will be the next logical step following Cluster, THEMIS and MMS for the very large and active European space plasmas community. Being one of the cornerstone missions of the future ESA Voyage 2035-2050 science program, it would further strengthen the European scientific and technical leadership in this important field.**

Baryonic matter in the Universe is almost exclusively a plasma. Energy conversion among electromagnetic, kinetic, thermal and non-thermal energies leads to the energization of particles in the plasma Universe. Examples of plasma regions of strong and sometimes spectacular particle energization are stellar coronae and winds, heliospheric and astrophysical shocks, planetary magnetospheres, supernova remnants, accretion disks and astrophysical jets. The physics of particle energization is a compelling science problem of major importance for the worldwide plasma community.

The key questions to be addressed covering the fundamental plasma processes that are responsible for most of particle energization in space plasmas are:

*(1) How are particles energized at shocks?*
*(2) How are particles energized during magnetic reconnection?*
*(3) How are particles energized by waves and turbulent fluctuations?*
*(4) How are particles energized in plasma jets?*
*(5) How are particles energized upon combination of different fundamental processes?*

In situ measurements are required to understand how particles are energized in space plasmas. In the solar system, the near-Earth space is the best laboratory for studying particle energization since very high resolution in situ measurements can be performed and transmitted to ground with high cadence. Furthermore, the near-Earth space provides a large range of different plasma conditions, which are ideal to explore all the fundamental energization processes. Due to similarities with other solar and astrophysical plasma regimes, many of the results obtained in near-Earth space can be exported to those environments where in situ measurements are not possible.

Over the last two decades, the multi-point constellations ESA/[Cluster](#) (Escoubet+, AG, 2001), NASA/[THEMIS](#) (Angelopoulos+, SSR, 2008) and NASA/[MMS](#) (Burch+, SSR, 2016) have significantly improved our understanding of particle energization. Also these missions have shifted the focus of the space plasma community towards quantitative fundamental plasma physics as demonstrated by the large number of publications in high impact journals (Nature, Science, Physical Review Letters, etc.) devoted to these topics.

The questions outlined above require resolving cross-scale coupling, nonlinearity and nonstationarity which cannot be done with existing multi-point constellations. These constellations were designed for and optimally studied one scale at a time. Moreover, 4-point measurements cannot resolve nonlinearity and nonstationarity. To overcome all these limitations, in situ measurements by a plasma observatory consisting of at least 7 spacecraft covering fluid, ion and electron scales are needed to fully answer these questions. Such a plasma observatory requires an ESA L-class mission. Different mission concepts can be envisioned. *Constellation of 7 spacecraft*: This would be an optimization of the ESA Cross-Scale concept. The constellation would consist of 7 spacecraft with identical platforms and possibly two different types of payload, one tailored for electron scales and the other for ion/fluid scales. The actual repartition between the two types of payload would depend on the technological and scientific developments. *Mother and 6 daughters*: This would be an optimization of the earlier JAXA-CAS SCOPE concept, a constellation of one mother spacecraft and 6 identical daughters all carrying identical scientific payload. The mother spacecraft will have a SCOPE-like or THOR-like high-resolution payload while the 6 smaller daughters will have lower resolution payloads.

A number of technological developments, some of them already ongoing, on platforms, payload and operations, will allow to reduce the complexity and cost of such plasma observatory. In addition to these developments, the currently strong worldwide interest in such constellations, e.g. in the USA, China, Japan, Russia, would enable significant international participation to this observatory.

## 2 Motivation

Conversion among electromagnetic, kinetic, thermal and non-thermal energies in collisionless astrophysical plasmas results in the energization of electrons, protons and heavier ions. A complete comprehension of the energization mechanisms is far from being achieved.

The near-Earth space (Figure 1) is the best natural laboratory to study particle energization since very high resolution in situ measurements are possible therein. These measurements are required to identify the exact energization mechanisms and how they depend on different plasma conditions, which are found in different regions such as the pristine solar wind, the foreshock, the bow shock, magnetosheath (shocked solar wind), the magnetopause and magnetotail current sheet and jet braking region. These conditions in dimensionless parameter space are similar to those in solar and astrophysical plasmas, so that much of the understanding of particle energization processes from near-Earth space can be exported there.

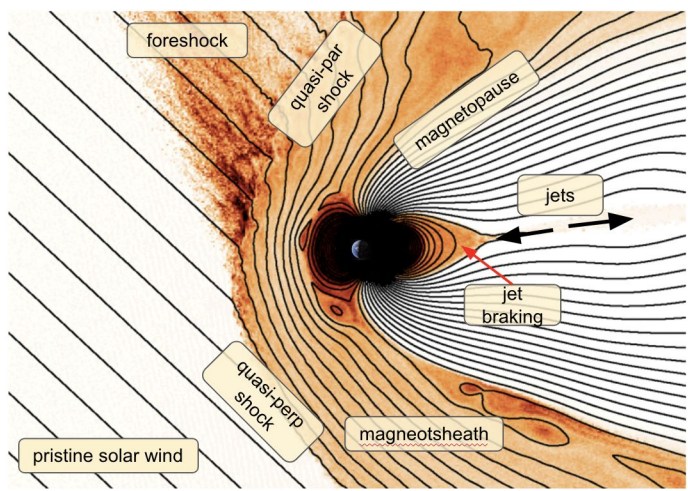

*Figure 1. Near-Earth regions important for particle energization. Color scale is ion temperature in 2D global hybrid simulation, courtesy of D. Krauss-Varban.*

Following the pioneering work in the 60's with the first satellites carrying in situ plasma instrumentation in near-Earth space (e.g. Luna1-3, Mariner-2, Explorer1&3), a large number of single spacecraft missions have explored the near-Earth space providing an increasing number of details of plasma regions (e.g. Helios, ISEE, AMPTE, Geotail, ACE, Wind, Interball and Equator-S) and a first framework to understand particle energization. In the last two decades, multi-point in situ measurements from Cluster, THEMIS and MMS with much improved instrument resolution and data quality, together with boosting computational capabilities of numerical simulations, have allowed major advances. These lead to a natural evolution from a qualitative description of phenomena to a quantitative study of fundamental plasma processes. Following this evolution, a very large fraction of scientists have made major efforts to improve our understanding of particle energization produced by shocks, magnetic reconnection, turbulence and plasma jets, and this theme has emerged as a compelling science topic that needs to be studied in the coming decades.

Despite all these results, the multi-spacecraft missions have clearly highlighted many critical limitations which hinder further significant advances and to discover new physics, and eventually reach closure on the compelling theme of particle energization. Two blocking points are present: the limited number of observation points (number of spacecraft) and the limited resolution of current instrumentation (both in time and in velocity space). As for the number of points available, recent in situ observations and simulations unequivocally demonstrate that particle energization mechanisms are strongly coupled on three fundamental scales simultaneously: the electron-kinetic, ion-kinetic, and fluid scales, while Cluster and MMS can only address one scale at a time, fluid or ions and ions or electron, respectively. Current observations also show that most of the plasma structures which are responsible for particle energization are 3D, non-linear and non-stationary. Yet, the existing 4-spacecraft missions, i.e. Cluster and MMS, can only resolve linear and stationary structures. As for the resolution of in situ instrumentation, many limitations need to be overcome too. The MMS mission has made substantial improvements with respect to Cluster, namely in the measurements of ion and electron 3D distribution functions at their characteristic scales. Nevertheless, substantial further improvements are still needed, namely high-time resolution measurements of mass-resolved ions, high angular and energy resolution of solar wind ions and improved sensitivity and accuracy of electric and magnetic field measurements. To overcome all these limitations, at least 7 points of measurements with high-resolution are needed, as will be illustrated by the examples in Section 3.

While these 7-point measurements are not yet available, very limited multi-scale measurements have been available after modification of the Cluster constellation starting from 2007. As an example, two spacecrafts have been separated in the magnetotail by few 10s of km (sub-ion to ion scales) while separated from the others by several 1000s of km (fluid scales). Results based on these observations have provided only limited insights on cross-scale coupling (Nakamura+, AnGeo, 2009; Zieger+, GRL, 2011). Another way to obtain multi-scale observations is to use conjunctions between existing constellations, e.g. between Cluster and MMS and between THEMIS and MMS. These fortuitous few conjunctions, while useful to address how small-scale processes are related over global scales under the same external conditions, cannot address the coupling between fluid, ion and electron scales in the same region of space.

A number of attempts have been made in the last decade to provide the community with a multi-point, multi-scale plasma observatory. The mission concept [Cross-Scale](Cross-Scale) (Schwartz+, ExpAstron, 2009) was selected by ESA for a competitive Phase-A study as Cosmic Vision M1M2 candidate and has been a very important step in this direction. The focus of Cross-Scale, a constellation of 7 identical spacecraft flying in formation and covering simultaneously two scales, is on the science of fundamental energization processes shocks, reconnection and turbulence. The JAXA [SCOPE](SCOPE) mission (Fujimoto+, 2009) is a constellation of 5 spacecraft addressing similar questions to Cross-Scale, with a couple of mother-daughter spacecraft measuring the microscopic scales and 3 additional spacecraft the macroscopic scales. The [EIDOSCOPE](EIDOSCOPE) (Vaivads+, ExpAstron, 2012) mission concept was proposed in the framework of an ESA M3 mission to complement the SCOPE constellation with one spacecraft focusing on particle acceleration. The mission concept [THOR](THOR) (Vaivads+, JPP, 2016) performed a competitive Phase-A study at ESA as an M4 candidate. It was one spacecraft carrying the highest resolution in situ payload ever conceived, addressing the energization of particles by turbulence. The mission concept [PROSPERO](PROSPERO), submitted as an ESA F1 candidate, was a constellation of 8 cubesats and one mother spacecraft addressing the structure of particle energization sites in near-Earth space. Finally the mission concept [Debye](Debye), also submitted as an ESA F1 candidate, was a mother spacecraft accompanied by three smaller daughters to study electron heating in plasma.

All these efforts clearly demonstrate the strong need of the space plasmas community for a new plasma observatory consisting of a 7-points constellation providing high-resolution measurements in near-Earth space. The ESA Cosmic Vision already includes a magnetospheric swarm as one of the mission scenarios identified to address the Theme 2.1 - *"How does the Solar System work? From the Sun to the edges of the Solar System"*. Such observatory corresponds to an L-class mission and is the next logical step for the world-leading European space plasmas community. It would also provide very valuable input and synergies to the planetary, solar and astrophysical communities

## 3 Science theme and questions

The science theme that this White Paper addresses is: **"How are charged particles energized in space plasmas"**. This is a theme of pivotal importance, as demonstrated by the very large number of refereed publications in high impact journals (Nature, Science, Physical Review Letters, etc.), dedicated books, review articles and special issues. Numerous international research collaborations have been devoted to this topic, such as workshops, working groups and ISSI forums and ISSI teams (see list in the bibliography).

As anticipated in the previous section, energization of particles in space plasmas, including solar and astrophysical plasmas, is related to fundamental plasma processes such as shocks, magnetic reconnection, turbulence and waves, plasma jets (Aschwanden, 2005; Balogh+, SSR, 2013; Ji+, Science, 2015). However, the exact physical mechanisms of energization behind those processes, as well as coupling between the different mechanisms, are not understood quantitatively and therefore key science questions to be answered by the plasma observatory are:

1. *How are particles energized at shocks?*
2. *How are particles energized during magnetic reconnection?*
3. *How are particles energized by waves and turbulent fluctuations?*
4. *How are particles energized in plasma jets?*
5. *How are particles energized upon combination of different fundamental processes?*

Particle energization for all the above questions involves simultaneous coupling of processes at electron, ion, and fluid scales. In addition to this scale coupling, even at a given scale, plasma energization structures are highly nonlinear and nonstationary. Resolving the complex scale coupling, as well as nonlinearity and nonstationarity, is required to reach closure on these questions. All this is not possible with current four-point observations, and measurements by the new plasma observatory are required.

For each of the five questions above, we provide several examples illustrating the need for such new measurements. The plasma observatory shall probe different near-Earth space regions, such as the pristine and shocked solar wind, the bow shock, magnetopause, and the magnetotail current sheet, which are all representative of solar and astrophysical plasma environments.

## *3.1 How are charged particles energized at shocks?*

Plasma shocks are formed when supersonic plasma flows encounter obstacles. The transition from supersonic to subsonic flow leads to the deceleration, compression and heating of the incident plasma and particle acceleration. A large fraction of particle heating and acceleration occurring in space and astrophysical plasmas, is believed to be produced at collisionless shocks (e.g. Jones+, SSR, 1991). Important examples are shocks generated by supernovae, in galaxy clusters, and stellar wind generated interplanetary, planetary and termination shocks.

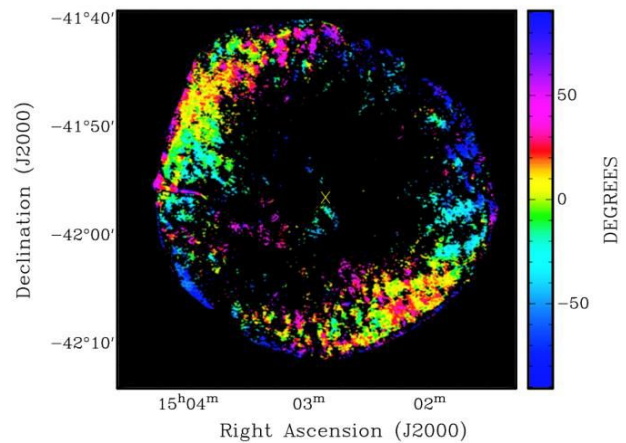

*Figure 2. Strong particle acceleration at supernova remnant shock SN 1006. Magnetic field is radial in yellow regions (quasi-parallel) and perpendicular to radial direction in blue regions (quasi-perpendicular). The most efficient particle acceleration is attained at quasi-parallel portion of the shock (Reynoso+, AJ, 2013).*

The terrestrial bow shock, generated by the interaction of the solar wind with the Earth's magnetosphere, is the most studied shock using in situ data. Key parameters controlling particle energization at shocks are the Mach number and the shock angle, which is the angle between the shock normal and inflowing magnetic field. Astrophysical shocks are much larger and can have significantly higher Mach numbers but the physical mechanisms of particle acceleration can be very similar. This makes the terrestrial bow shock the best natural laboratory to study particle energization at shocks, allowing much of the knowledge to be exported to distant astrophysical shocks where in situ measurements are not possible.

Collisionless shocks are inherently multi-scale processes. The shock transition occurs due to highly non-equilibrium physics coupling electron, ion and fluid scales and creates very complex plasma structures at each of these scales and these structures are important for particle energization.

### 3.1.1 Ion injection to suprathermal energies

The Diffusive Shock Acceleration (DSA) mechanism (Bell+, MNRAS, 1978) is a well-established mechanism to produce energetic particles at collisionless shock waves. In order for DSA to proceed efficiently, however, a fraction of the seed population particles needs first to be pre-accelerated to suprathermal energies, a process called injection. Despite its importance, a complete understanding of the physical mechanisms of injection is far from being achieved. Without this understanding, it is not possible to establish a meaningful injection model, which is crucial to interpret remote observations at astrophysical shocks. Numerical simulations of the shock structure and ion dynamics have attempted to obtain injection models without free parameters. Kucharek et al., (1991) proposed a two step process where an initial acceleration of ions in the shock ramp is followed by multiple ion reflections across the shock. Figure 3a shows an example of kinetic simulations of a quasi-parallel shock (Caprioli+, ApJL, 2015). The shock discontinuity evolves on kinetic timescales (a few proton gyro times) leading to a formation of nonlinear and nonstationary upstream structures at kinetic scales. Ion injection is produced by reflection and scattering of incoming ions with such structures and depends on ion masses and energies. Therefore accurate measurements resolving such kinetic structures and particle distributions at multiple points are crucial for understanding how the injection actually works.

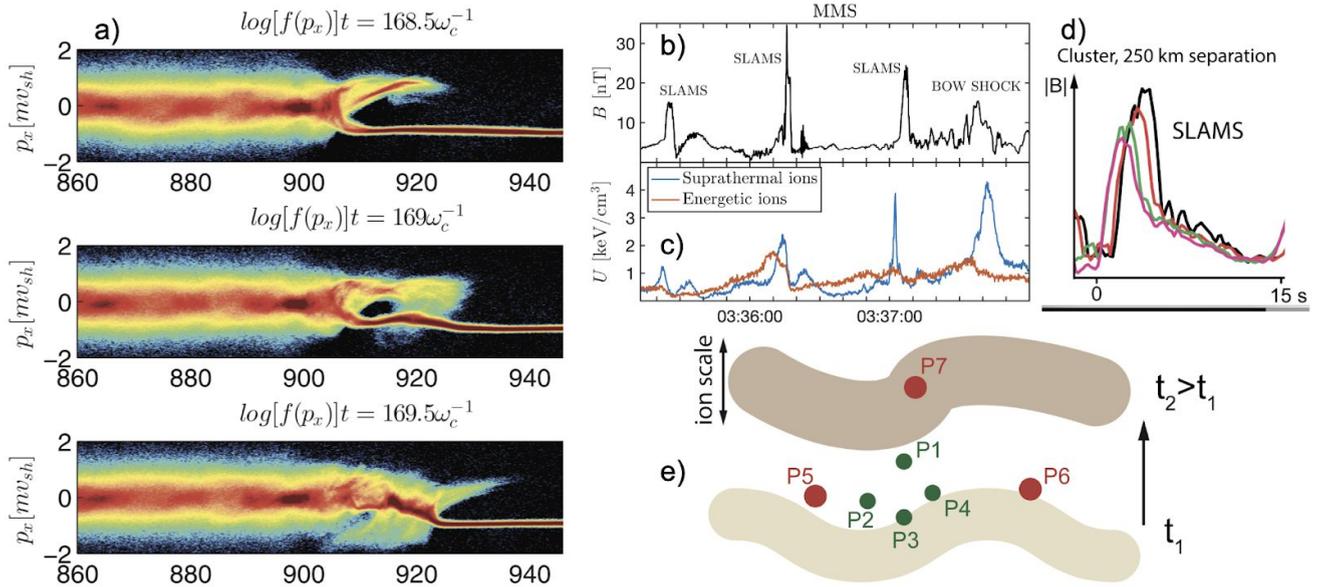

*Figure 3. Quasi-parallel shock. a) Kinetic simulations of the time evolution of ion injection at a quasi parallel shock with Mach number M=20 (Caprioli+, ApJL, 2015). b-c) MMS measurements of ion energization by SLAMS (Johlander+, 2019) d) SLAMS evolution in Cluster data (Lucek+, JGR, 2008), e) sketch illustrating the need of 7 spacecraft to resolve spatio-temporal variations at SLAMS.*

An important example of upstream structures are the Short Large Amplitude Magnetic Structures (SLAMS), which are common features of quasi-parallel shocks (Schwartz+, GRL, 1991; Lucek+, JGR, 2008) and are very efficient for ion scattering and energization, see Figure 4b,c (Behlke+, GRL, 2003; Johlander+, ApJL, 2016; Johlander+, 2019). Figure 3d shows Cluster 4-point measurements of one SLAMS demonstrating that they are neither linear (planar) nor stationary since the amplitude of the SLAMS appears to grow as it propagates. Cluster and MMS measurements cannot distinguish spatial from temporal variations.

At least 7 measurement points are needed to fully characterize the spatial structure and separate the growth of SLAMS from their motion. One point in addition to 4 spacecraft will allow the growth in time of the SLAMS amplitude to be resolved along the propagation direction (P7 in Fig 3e) and address nonstationarity. Having two additional spacecraft in different directions perpendicular to the propagation direction (P5 and P6) will resolve the 3D morphology of SLAMS and address nonlinearity. Both magnetic and electric fields and 3D ion distribution functions, both in thermal and suprathermal ranges, shall be measured at ion scales (typical cadence ~ 0.1 s). Ion measurements shall also resolve mass composition, at least protons and alpha particles since the different ions have different gyroradii and interact differently with SLAMS.

Such measurements will allow to assess the efficiency of solar wind ion energization due to SLAMS, as well as hot flow anomalies (Schwartz+, JGR, 2018) and other foreshock structures (Liu+, SciAdv, 2019). Composition measurements will be crucial for understanding preferential energization and the energy partition among species, something of key importance for cosmic rays (Meyer+, Nature, 1978; Miceli+, Nature, 2019).

### 3.1.2 Electron heating

Another very important question illustrating the need of new multi-point, multi-scale measurements at shocks is how electrons are heated in the shock transition (Treumann+, AAR, 2009). Early in-situ observations have shown that a substantial part of incident ion ram kinetic energy is converted into electron thermal energy (Schwartz+, JGR, 1988), raising the fundamental question of which mechanisms are responsible for this heating. Adiabatic heating is a natural candidate yet a systematic deviation from adiabaticity is observed and is not yet understood (Schwartz+, JGR, 1988; Wilson+, JGR, 2014; Chen+, PRL, 2018).

Limited Cluster measurements at electron scales were possible in one point only and have hinted that heating scales could be as small as a few electron scales (Schwartz+, PRL, 2011). Recent MMS four-point measurements at electron scales (Chen+, PRL, 2018) have been able to resolve electron scale heating and strongly suggested that the heating process is highly nonadiabatic, far beyond the simple picture of a quasistatic cross-shock potential. Yet no simultaneous measurements at ion scales were available with MMS. Such simultaneous measurements are needed since shock non-stationarity can strongly affect electron heating and typically occurs at both ion scales (Johlander+, ApJL, 2016) and electron scales (Dimmock+, SciAdv, 2019). Additionally, trapping within shock ripples and other electron-scale structures can allow electrons to reside close to the shock ramp where they can gain even more energy from the shock electric fields.

At least 7 measurement points are needed, either simultaneously at both electron and ion scales to resolve scale coupling, or all at electron scales to resolve nonlinearity and non-stationarity of heating structures. These measurements will allow a consistent model of electron heating at the shock and estimates of quantities such as heating rates to be obtained, which could be used to help interpreting X-ray emissions from solar and astrophysical plasmas, such as e.g. those from galaxy clusters where shock heating is suggested to be important (McNamara+, Nature, 2005).

## *3.2 How are charged particles energized during magnetic reconnection ?*

A large fraction of particle energization occurring in astrophysical plasmas is produced at current sheets, which separate magnetic fields and plasmas of different types. The dominant process of particle energization in current sheets is magnetic reconnection, a fundamental process which converts magnetic energy into kinetic, thermal and non-thermal energies of plasmas (Yamada+, Rev. Mod. Phys., 2010). Energization by reconnection plays a key role in the heating of stellar coronae (Cassak+, ApJ, 2008) and accretion discs (Goodman+, ApJ, 2008), in driving their supersonic winds, in relativistic jets from black holes and other compact objects, in powering giant radio galaxies (Kirk+, ApJ, 2003) and in accelerating cosmic rays (Lazarian+, ApJ, 2009; Drake+, ApJ, 2010). In the near-Earth space, particle energization by reconnection is the main process responsible for the transfer of energy from the solar wind into the magnetosphere and it occurs at the magnetopause and magnetotail current sheets.

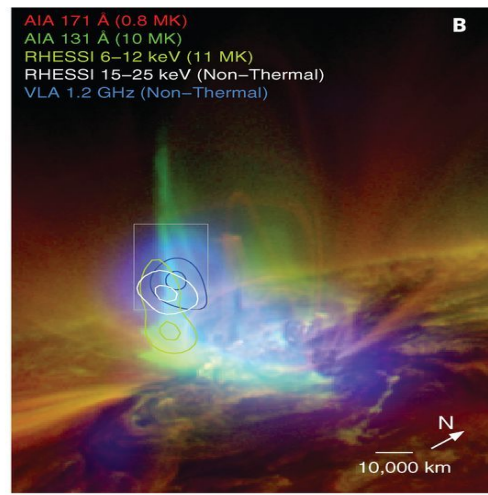

*Figure 4. Radiation emitted by energized particles in a solar flare. A radio source (blue) is observed at the top of hot flaring loops (~10 MK), which is nearly cospatial with a nonthermal HXR source (white contours) seen by the RHESSI spacecraft. Adopted from Chen+, Science, 2015*

Magnetic reconnection and associated particle energization are inherently multi-scale processes. Reconnection occurs in thin current sheets where kinetic effects first decouple ions from the magnetic field in the ion-scale diffusion region (Vaivads+, PRL, 2004) and then electrons in the electron-scale diffusion region (Burch+, Science, 2016; Torbert+, Science, 2018). Microscale processes occurring at these kinetic scales control particle energization, which eventually affect much larger volumes at fluid scales and beyond. On the other hand, large-scale processes control the location and formation of thin current sheets, and thus directly affect how reconnection initiates and evolves. It is therefore essential to simultaneously measure both the large-scale and the kinetic-scale plasma processes. In addition to this scale coupling, simulations and in situ observations indicate that, even at a given scale, the structure of reconnection regions, can be rather nonlinear and nonstationary (Lapenta+, NatPhys, 2015; Price+, GRL, 2016; Phan+, GRL, 2016; Swisdak+, GRL, 2018; Pucci+, ApJ, 2017; Cozzani+, PRE, 2019).

### 3.2.1 Electron heating

One science case demonstrating the need for new multi-scale measurements is electron heating in reconnection. Different mechanisms have been proposed, such as heating due to electric potential drops aligned to the background magnetic field or heating by different plasma waves such as whistlers or electrostatic waves (Egedal+, NatPhys, 2012; Muñoz+, PRE, 2018; Che+, Nature, 2011). Figure 5a shows an example from asymmetric reconnection for which 3D simulations predict efficient parallel electron heating, while no such heating is observed in 2D simulations (Le+, PoP, 2018) indicating that 3D effects are crucial. Recent MMS observations at the magnetopause in Figure 5b have established that heating is due to parallel potential drop (Graham+, GRL, 2016), however the formation mechanism and the location of the heating regions is still not understood. Simulations show that these 3D heating regions are formed at sub-ion scales. On the other hand, the reconnection process itself is driven over a volume corresponding to many ion scales. Understanding the coupling between these two scales is therefore crucial to understand electron heating. Neither Cluster nor MMS 4-point measurements cannot simultaneously address 3D structures at both scales.

At least 7 measurement points are needed. In addition to 4 spacecraft at sub-ion scales, 3 additional spacecraft separated by many ion scales are required to simultaneously observe the inflow (P5 and P6 in Fig. 5a) and outflow regions (P7), e.g. to measure the boundary conditions and to estimate the geometry of the reconnection site. Magnetic and electric fields and electron distribution functions shall be measured in 3D at least on few spacecraft at sub-ion scales (typical cadence tens of ms).

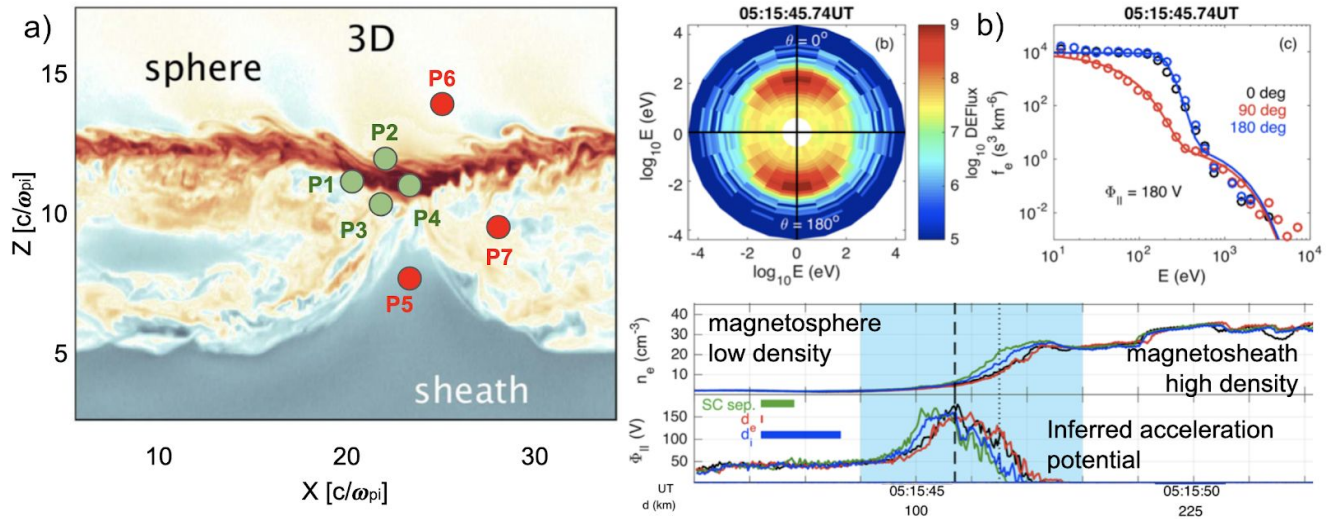

*Figure 5. Electron heating parallel to ambient magnetic field in asymmetric reconnection. a) Numerical simulations show efficient parallel heating in 3D simulations (Le+, PoP, 2018). b) MMS small-scale observations show similar heating (top panels) at low density side and it is consistent with being due to parallel electric field potential (bottom panels) (Graham+, GRL, 2016). Schematic 7 spacecraft constellation, P1-P7, would allow to address the role of cross-scale coupling in the formation of acceleration sites.*

In addition to heating due to parallel potential drops, plasma waves can also heat electrons. However, the relative importance between wave and potential drop heating is not understood and is an open question of fundamental importance. As an example, electrostatic waves (EWs) and electrostatic solitary waves (ESWs) are observed together with electron heating at reconnection separatrices (Cattell+, JGR, 2005; Viberg+, GRL, 2013) but their role for electron heating is not established. Recent MMS observations allowed for improved measurements of EW and ESW properties compared to Cluster (Holmes+, JGR, 2018; Tong+, GRL, 2018, Steinvall+, GRL, 2019) and clearly indicated that they are nonlinear and often nonstationary structures whose spatial and temporal evolution cannot be resolved with 4-point measurements.

At least 7 measurement points are needed to fully characterise these electrostatic structures and to establish their actual role in electron heating. All these measurements will allow the relative importance of potential jumps and wave-particle interactions for electron heating to be quantitatively assessed, which is a topic of general importance for other heliospheric and astrophysical plasmas (Zimbardo+, PRL, 2013).

### 3.2.2 Differential ion energization

Another very important science problem is how different ion species are energized during reconnection. A typical example can be found in the Earth's magnetotail where oxygen ions can have significant concentrations (Kistler+, JGR, 2005). Due to their different gyroradii, heavier ions decouple from the magnetic fields at larger scales than protons, substantially modifying the structure of the reconnection diffusion region where ions are energized. This can lead to important effects on ion energization such as reducing the speed of reconnection jets and the reconnection rate (Shay+, PRL, 2004). Another possible effect is reducing the number of secondary magnetic islands and slowing down their merging process (Karimabadi+, PSS, 2011), thus leading to a less efficient energization by these structures (Pritchett+, PoP, 2008; Oka+, PRL, 2008). At present, only very limited observations of simultaneous proton and oxygen energization in the diffusion region exist by Cluster spacecraft (Wygant+, JGR, 2005).

At least 7 spacecraft observations are needed to resolve simultaneously both proton and oxygen scales. High time resolution measurements of mass-resolved ion distribution functions (cadence from several hundreds of ms to 1s) are also needed since they are not available on either Cluster (4s cadence) or MMS (10s cadence).

These observations will be very important to understand particle energization during reconnection when the concentration of heavy ion is high, such as in reconnection during storm times in the terrestrial magnetosphere (Kistler+, JGR, 2010), at the magnetopause of Ganymede (Collinson+, GRL, 2018) or in impulsive solar flares (Drake+, ApJ, 2009; Kumar+, ApJ, 2017).

### *3.3 How are particles energized by waves and turbulent fluctuations?*

Waves and turbulent fluctuations are ubiquitous in space and astrophysical plasma. Significant particle energization is related to the dissipation of different types of plasma waves and turbulent fluctuations, such as Kelvin-Helmholtz, kinetic Alfvén, whistlers, different kind of solitary waves and coherent structures. Examples can be found in galaxies (Hajivassiliou, Nature, 1992; Schekochihin+, PoP, 2006), stellar interiors (Brandenburg+, RPP, 2011), interstellar (Elmegreen+, ARAA, 2004; Arzoumanian, A&A, 2011) and interplanetary (Bruno, LRSP, 2013; Alexandrova+, SRS, 2013) media and planetary magnetospheres (Uritsky+, JGR ,2011; von Papen+, JGR, 2011).

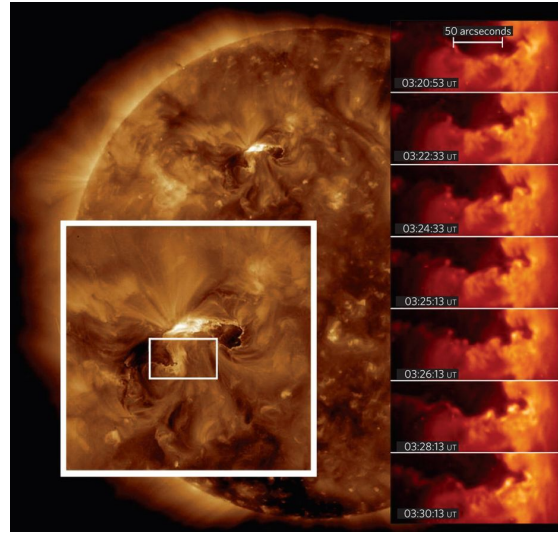

*Figure 6. CME eruption seen by the SDO spacecraft (Ofman+, ApJ, 2011). The large white box is a magnified view of the erupting structure. Right: temporal evolution of Kelvin–Helmholtz vortices at the boundary between the dark region, corresponding to evacuated material, and the surrounding ambient material (seen in the small white box). Strong heating may occur within Kelvin-Helmoltz vortices through wave-particle interaction.*

Near-Earth turbulent regions, such as the pristine and shocked solar wind (magnetosheath) and magnetopause and magnetotail boundary layers, allow particle energization mechanisms by different types of waves and turbulent fluctuations to be observed under different conditions, over a broad range of scales. Due to similarities with other solar and astrophysical regimes, many of the results obtained in near-Earth space are helpful for understanding other plasma environments. As an example, Kelvin-Helmotz waves and turbulence are observed both in the solar corona (Figure 6) and in the terrestrial magnetopause boundary.

Such processes are intrinsically connected across multiple scales. In turbulence, for example, the energy of large fluid-scale structures is transported towards smaller scales through a turbulent cascade of nonlinear interactions (Bruno+, LRSP, 2013), making it crucial to resolve scale-coupling through multi-points measurements (Matthaeus+, 2018). When the energy reaches the ion and electron scales, plasma kinetic processes arise, such as nonlinear damping of waves and dissipation in coherent structures which convert the energy of the turbulent fluctuations into plasma heating and particle acceleration (Salem+, ApJL, 2012; Chen, JPP, 2016; Pezzi+, PRE, 2017). The energy transfer and dissipation processes, the scale at which each processes occur, the energy partition between protons, heavier ions or electrons are still open questions. The increasing performance of numerical simulations (Karimabadi+, PoP, 2013; Valentini+, NJP, 2016; Franci+, ApJ, 2018) and theoretical efforts (Schekochihin+, JPP, 2016; Servidio+, PRL, 2017) allow these questions to be addressed. However, currently available multi-point in situ measurements clearly demonstrate the need for a new dedicated mission that can address the cross-scale coupling and nonlinearity (Chasapis+, ApJ, 2018; Sorriso-Valvo+, PRL, 2019; Chen+, NatComm, 2019).

### 3.3.1 Particle energization at coherent structures

One very important science case demonstrating the need for new multi-scale measurements is particle energization due to the energy dissipation in coherent structures generated by turbulence, which are localized both in space and time (Matthaeus+, PTRSA, 2015). These include thin current sheets, magnetic islands, isolated flux tubes, and small-scale vortices. Figure 7 shows prediction from numerical simulations and in situ MMS observations data which confirm that strong energy dissipation and particle energization occurs in kinetic-scale regions which are associated with strong electric currents (Wan+, PoP, 2016; Chasapis+, ApJ, 2018). These are also regions where non-Maxwellian features of particle distribution functions are observed. Cluster measurements at ion scales have shown energy dissipation and particle energization at thin current sheets observed in the turbulent pristine solar wind and magnetosheath (Sundkvist+, Nature, 2005; Sundkvist+, PRL, 2007; Chian+, ApJL, 2011; Perri+, PRL, 2012; Chasapis+, ApJL, 2015; Chasapis+, ApJ, 2017), which can be associated to small-scale reconnection (Retino+, NatPhys, 2007) predicted by Matthaeus+, Phys. Fluids, 1986. However, Cluster measurements could not resolve the electron scales. Recent MMS measurements have resolved electron-scale coherent structures. As an example, electron-scale reconnection events are observed in the turbulent magnetosheath (Phan+, Nature, 2018; Stawarz+, ApJ, 2019) and are associated with dissipation at electron scales. MMS has also observed other electron-scale coherent structures, such as magnetic holes and vortexes which are associated with strong electron energization (Huang+, JGR, 2017). MMS, on the other hand, cannot provide simultaneous observations at ion and fluid scales that are driving the turbulent energy input and coherent structure formation and therefore does not allow the coupling between scales to be addressed.

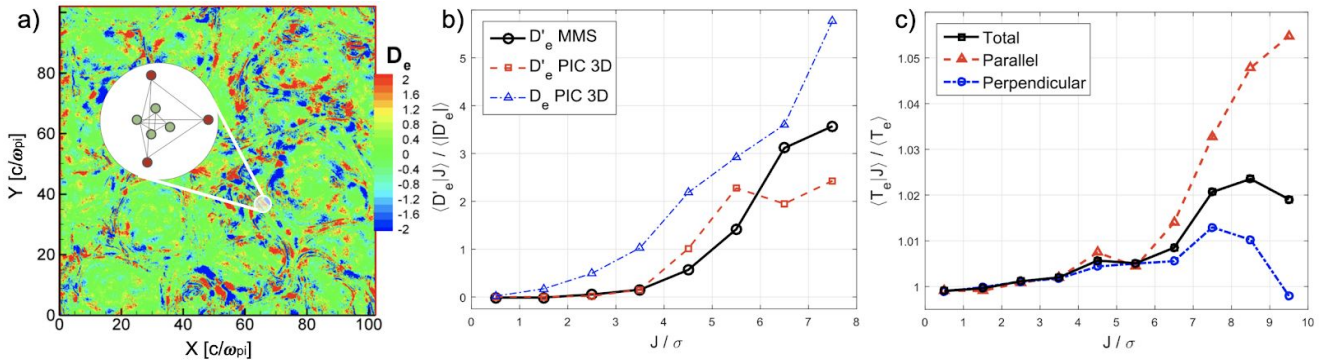

*Figure 7. Energy dissipation and particle energization in turbulence due to coherent structures. a) spatial distribution of energy dissipation as seen in PIC simulations, showing that dissipation is concentrated at kinetic-scale coherent structures (Wan+, PoP, 2016). b) simulations and MMS observations (Chasapis+, ApJ, 2018) showing that the strongest dissipation occurs in regions of highest current. c) MMS measurements of electron heating showing that strong parallel electron heating is associated to dissipation and high currents.*

At least 7 measurement points distributed in space to cover multiple scales are needed to resolve scale coupling in turbulent coherent structures and assess how it controls particle energization mechanisms. This would allow the correct identification and full description of coherent structures at different scales, as well as the turbulence conditions on larger scales. It would remove the severe approximations arising in the 4-spacecraft techniques, which are mostly based on linearity and stationarity assumptions for obtaining 3D propagation and shape. In addition, measurements of particle distributions should be improved with respect to MMS e.g. some electron-scale spacecraft should have higher time resolution electron measurements and higher phase-space resolution to resolve the fine details of the particle distributions that can reveal the nature of the dissipative processes (Schekochihin+, JPP, 2016; Servidio+, PRL, 2017). High time resolution mass-resolved ions should also be measured to evaluate the differential energization of protons and α particles (Perrone+, ApJ, 2013).

Future multi-point observations of particle energization in coherent structures can be important to help interpreting remote observations e.g. from the interstellar medium where large-scale turbulence properties can be measured (Armstrong+, ApJ, 1995) but no information is available on the smaller scales.

### 3.3.2 Particle energization by Kelvin-Helmholtz waves

Another science case demonstrating the need of new multi-scale measurements is particle energization occurring at Kelvin-Helmholtz (KH) waves and large amplitude vortices that develop during the turbulent stage of the instability (Karimabadi+, PoP, 2013). Kelvin-Helmholtz waves are generated in boundary layers with a flow shear between adjacent plasmas, such as at the terrestrial magnetopause (De Keyser+, PSS, 2003; Taylor+, AnGeo, 2012; Retino+, NatPhys, 2016). Kelvin-Helmholtz waves occurs also in other astrophysical plasmas, e.g. in the solar corona (e.g. Ofman+, ApJ, 2011), astrophysical jets (e.g. Bahcall+, ApJL, 1995), or even at the level of galaxy clusters (e.g. Walker et al. 2017). Energy conversion and particle energization during KH instability is an inherently multi-scale process. Recent supercomputer simulations predict that linear KH waves excited at the dayside magnetopause evolve into large-scale vortices eventually reaching a very turbulent stage where sites of substantial particle energization are produced at kinetic scales e.g. by secondary instabilities in vortices or small-scale reconnection (Rossi+, PoP, 2015; Nakamura+, NatComm, 2017). Cluster observations at fluid scales could only infer the existence of large-scale vortices at the magnetopause under 2D and steady-state assumptions (Hasegawa+, Nature, 2004). Further Cluster observations at ion scales have allowed to identify ion energization in KH vortices due to small-scale reconnection (Hasegawa+, JGR, 2009) or ion-scale magnetosonic waves (Moore+, NatPhys, 2016). More recent MMS observations have shown evidence of energization of both ions (Sorriso-Valvo+, PRL, 2019) and electrons (Eriksson+, GRL, 2016) in the turbulent phase of KH instability (Stawarz+, JGR, 2016). In order to address the scale coupling, simultaneous observations at least at two scales, fluid/ion or ion/electron, are needed in order to follow the overall vortex formation and evolution and resolve particle energization. Additionally KH vortexes can have a quite nonlinear and nonstationary structure, especially in the turbulent phase. Neither Cluster nor MMS 4-point measurements can provide such observations.

At least 7 measurement points are needed to resolve scale coupling of particle energization in KH instability and to resolve nonlinear and nonstationary structure of energization sites. These measurements could be used to interpret remote observations of KH-related particle energization, e.g. in CMEs (Foullon+, ApJL, 2011), solar (Li+, Nature, 2018) and astrophysical jets (Lobanov+, Science, 2001) and molecular clouds (Berné+, Nature, 2010).

### *3.4 How are particles energized in plasma jets?*

Jets are ubiquitous in astrophysical plasmas. They are observed in near-Earth space (Baumjohann+, JGR 1990; Phan+, Nature, 2000; Plaschke+, AnGeo 2013) and in the magnetospheres of other planets (Kasahara+, JGR 2013), in the solar corona (Innes+, Nature, 1997; McKenzie+, ApJ, 2009) and chromosphere (Shibata+, Science, 2007), as well as in other astrophysical objects such as jets from AGNs and protostars (Pudritz+ SSR, 2012).

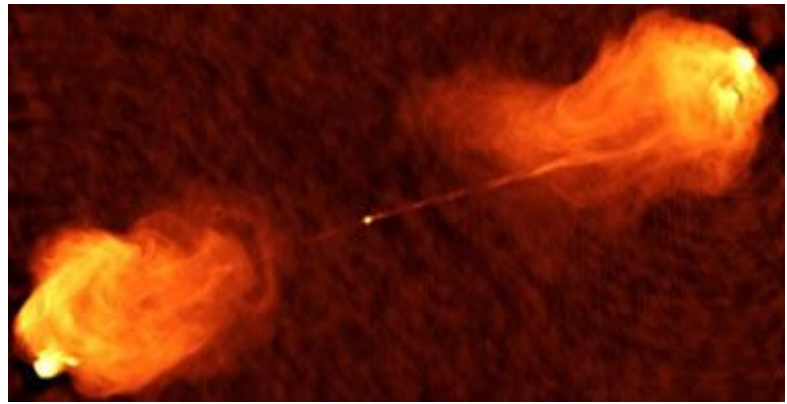

*Figure 8. A radio image of the galaxy Cygnus A showing the jet and radio lobes. The 'hot spots' that mark the shock fronts between the jet and the interstellar medium are clearly evident. Image: NRAO/AUI.*

Plasma jets are efficient for particle energization. In near-Earth space the most important examples are jets generated in the magnetopause and magnetotail current sheets as well as jets downstream of the shock. As jets propagate and interact with the ambient plasma, particles are heated and accelerated at the jet boundary, often referred to as plasma jet front (Ashour-Abdalla+, NatPhys, 2011; Khotyaintsev+, PRL, 2011; Zieger+, GRL, 2011; Fu+, NatPhys, 2013; Lapenta+, NatPhys, 2015; Liu+, Apj, 2019). Jets eventually stop in the jet braking regions upon interaction with obstacles, such as the Earth's magnetic dipolar field (Shiokawa+, GRL, 1997; Panov+, JGR, 2013), where particles can be efficiently accelerated (Vaivads+, AnGeo, 2011; Ukhorskiy+, JGR, 2018) and later injected into the inner magnetosphere (Sergeev+, GRL, 2009).

Particle energization at jet fronts and in braking regions involve a strong coupling of electron, ion and fluid scales. In the magnetotail, as an example, jet fronts have a large lateral extension at fluid scales (many Earth's radii) while having a much smaller thickness at kinetic scales (Runov+, JGR, 2011). Microscale processes occurring at fronts control the electric fields and the waves responsible for particle energization which, on the other hand, affect much larger volumes at fluid scales. It is therefore essential to simultaneously observe both the fluid-scale and the kinetic-scale plasma processes. In addition to this scale coupling, simulations and in situ observations indicate that, even at a given scale, the structure of fronts and braking regions can be nonlinear and nonstationary (e.g. Sitnov+, SSR, 2019).

### 3.4.1 Electron energization at jet fronts in the magnetotail

One science case demonstrating the need of new multi-scale measurements is electron energization at plasma jet fronts in the magnetotail. At large temporal and spatial scales, MHD simulations with test particles indicate that electron acceleration at jet fronts results from adiabatic betatron and Fermi mechanisms within large-scale magnetic flux tubes (Ashour-Abdalla+, NatPhys, 2011) and this prediction has been confirmed by observations (Fu+, GRL, 2011; Fu+, NatPhys, 2013), see Figure 9a,b. On the other hand, Cluster and THEMIS observations indicate that important conversion of electromagnetic energy occurs at kinetic scales (Angelopoulos+, Science, 2013), leading to strong electron energization by electric fields and waves at those scales (Khotyaintsev+, GRL, 2017). This was recently confirmed by MMS observations (Liu+, GRL, 2018). The scale coupling between fluid and kinetic scales is not understood and simultaneous observations at both scales are needed to obtain a full understanding of electron energization mechanisms. However, none of the available multi-point measurements can provide such observations. At least 7 measurement points are needed in order to resolve the coupling between fluid and kinetic scales.

Simulations and spacecraft observations also indicate that jet fronts are often very structured due to the development of different instabilities which range from electron-scale instabilities to hybrid-scale instabilities, such as the lower-hybrid drift instability, to ion-scale (e.g. kinetic interchange) and MHD instabilities, such as the interchange-ballooning, drift kink and Kelvin–Helmholtz instabilities (Divin+, JGR, 2015; Pan+, GRL, 2018; Sitnov+, GRL, 2018).

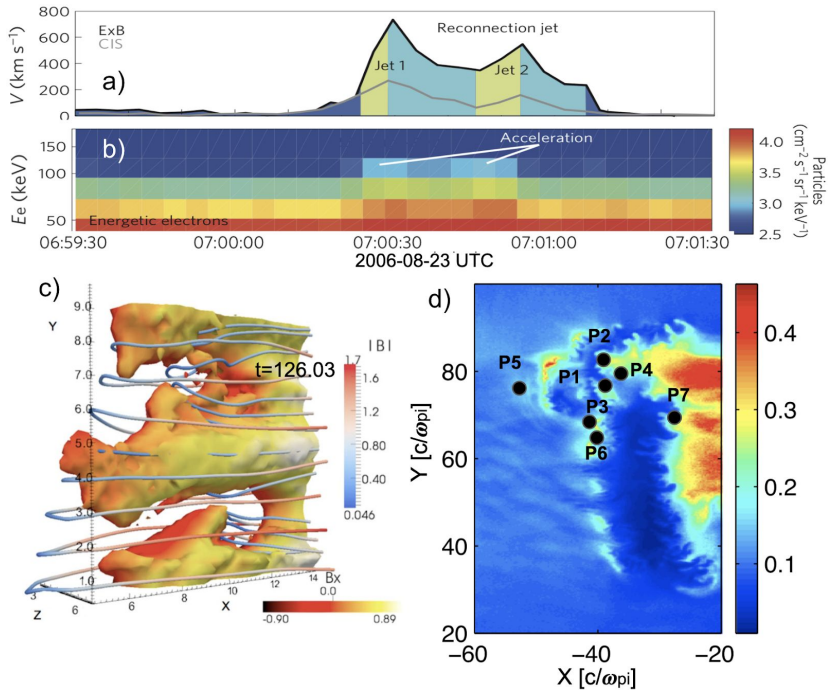

*Figure 9. Electron enerization at jet fronts. In situ observations of reconnection jets (a) associated to large-scale energization (b) (Fu+, NatPhys,2013). Numerical simulations showing nonlinear structures at jet fronts: c) 3D isosurfaces of density with superimposed field lines from MHD simulations (Lapenta+,GRL, 2011). d) parallel electron temperature from PIC simulations (Pritchett, JGR, 2016). Schematic 7 spacecraft constellation, P1-P7, would allow to characterise the spatial and temporal evolution of electron energization at jet fronts as well as scale coupling.*

Figure 9c shows a 3D MHD numerical simulations of a jet front undergoing interchange-like instability (Lapenta+, GRL, 2011). The front shows 3D nonlinear structures at fluid scales. Such structures also occur at kinetic scales and are important sites of electron energization, Figure 9d. Simulations also show the formation small-scale reconnection sites in the vicinity of the jet front, which can further contribute to electron heating (Lapenta+, NatPhys, 2015). All these structures are nonlinear and nonstationary structures whose spatial and temporal evolution cannot be resolved with 4-point measurements even at a given scale. At least 7 measurement points are needed to characterise the spatial and temporal evolution of jet fronts and associated electron energization sites.

Similar energization as that at magnetotail jet fronts is invoked for solar flares (Somov+, ApJ, 1997) where shock are also thought to be formed (Aurass+, ApJ, 2004). Future multi-point, multi-scale in situ observations in the magnetotail will therefore be very important to help understanding electron energization in the corona, for which only remote observations are possible.

### 3.4.2 Particle energization in magnetosheath jets

Another science case demonstrating the need of new multi-scale measurements is particle energization in magnetosheath jets. Magnetosheath jets are fluid-scale structures with a significantly enhanced dynamic pressure with respect to the ambient magnetosheath plasma and predominantly occur downstream of the quasi-parallel bow shock. Solar wind plasma can form jets as it passes inclined surfaces at shock ripples, resulting in an intrinsically nonlinear and nonstationary jet structure (Plaschke+, JGR, 2017). As jets propagate in the magnetosheath, they interact with and modify the ambient plasma (Plaschke+, AnGeo, 2018; Plaschke+, SSR, 2018). This interaction create sites where particles can be efficiently energized, such as magnetic bottle where Fermi acceleration operates (Liu+, ApJ, 2019) and kinetic-scale current sheets (Eriksson+, JGR, 2016) and shocks (Hietala+, AnGeo, 2012). All these sites have a nonlinear and nonstationary structure whose spatial and temporal evolution cannot be resolved with current 4-point measurements even at one given scale. Additionally, current 4-point measurements cannot capture the multi-scale nature of jets where fluid and kinetic scales are strongly coupled

At least 7 measurement points are needed to fully characterise the spatial and temporal evolution of magnetosheath jets and associated particle energization sites as well as scale coupling therein. Jets should be universally occurring downstream of collisionless shock such as other planetary or exoplanetary bow shocks or astrophysical shocks. Future multi-scale observations in the magnetosheath will be very important to help understanding particle energization by jets in those plasmas.

### *3.5 How are particles energized upon combination of different fundamental processes?*

One of the major insights coming from Cluster, THEMIS and MMS missions, as well as from recent 3D supercomputer simulations, is that in real systems different particle energization processes are often combined in a complex way. For example, there are observations of small-scale reconnection in turbulence (Retino+, NatPhys, 2007; Phan+, Nature, 2018), as well as turbulence in large-scale reconnection (Fu+, GRL, 2017), observations of turbulence at shocks (Schwartz+, GRL, 1991), as well as reconnection at shocks (Wang+, GRL, 2019; Gingell+., GRL, 2019). Large-scale kinetic simulations even show the combination of shocks, reconnection, turbulence and jets in the same region (Karimabadi+, PoP, 2014; Matsumoto+ Science, 2015). There are clear indications that the combination of different processes, each per se important for particle energization, can make energization even more efficient. Also scale coupling and the formation of nonlinear and nonstationary energization sites will be further enhanced during combination of processes, see Figure 10. All this presents a big challenge for theoretical models and in situ observations. Addressing these questions requires designing missions that can go beyond the approach of existing missions focusing on a single process at a time.

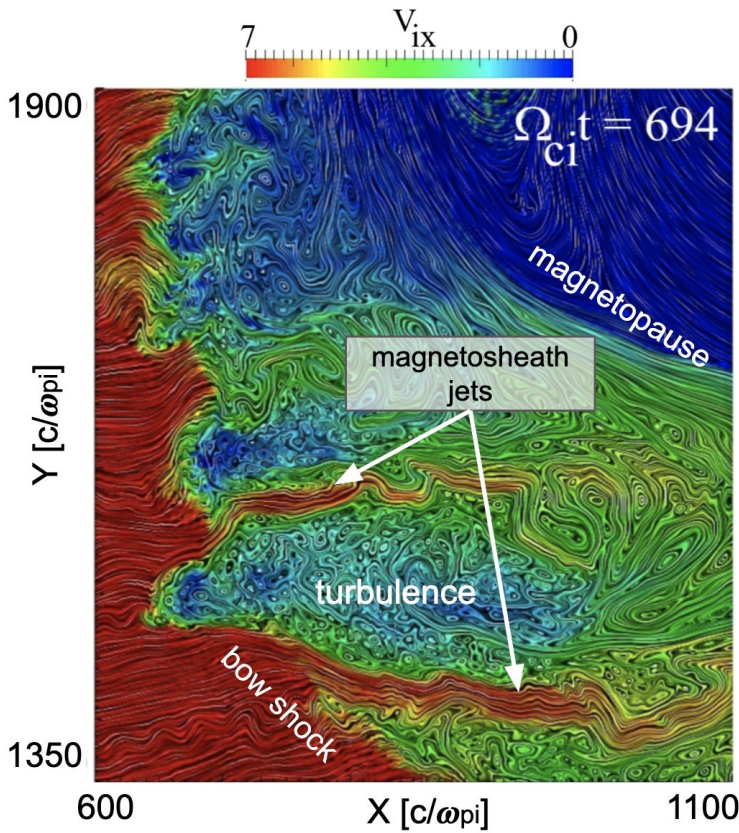

*Figure 10. Examples from numerical simulation where shocks, turbulence, reconnection and jets all couple together (Karimabadi+, PoP, 2014).*

### 3.5.1 Turbulent reconnection

One science case demonstrating the need of new multi-scale measurements is particle energization during turbulent reconnection, where reconnection and turbulence strongly couple (Lazarian+, ApJ, 1999; Daughton+, NatPhys, 2011). MMS measurements have shown examples of electron diffusion regions (EDR) that are laminar both at the magnetopause (Burch+, Science, 2016) and in the magnetotail (Torbert+, Science, 2018). Figure 11a-d presents one such laminar EDR at the magnetopause, initially reported by Chen+, GRL, 2016 and whose structure has been reconstructed with the FOTE method (Fu+, JGR, 2015). This method assumes linear variations of the magnetic field in the volume surrounding MMS and has allowed the magnetic topology of the EDR to be reproduced, Figure 11c, together with a map of electron energization, Figure 11d (Fu+, 2019). For such laminar EDRs, 4-point measurements can reveal the electron and ion dynamics in a satisfactory manner. However, simulations (Daugthon+, NatPhys, 2011; Lapenta+, NatPhys, 2015) and spacecraft observations (Phan+, GRL, 2016; Fu+, GRL, 2017; Cozzani+, PRE, 2019) indicate that the reconnection diffusion regions can be rather turbulent, with the formation of many intermittent structures such as thin current sheets, magnetic islands, vortexes and magnetic holes which are very efficient to energize particles (Retino+, JGR, 2008; Chen+, NatPhys, 2008; Wang+, NatPhys, 2016; Huang+, ApJ, 2018). Figure 11e-g shows an example of such turbulent reconnection in the Earth's magnetotail. The four Cluster spacecraft crossed the ion diffusion region (IDR) and observed strong turbulence of magnetic fields characterized by intense current filaments, Figure 11e, associated with strong energy dissipation, Figure 11f. Current filaments occur mostly at magnetic nulls (Xiao+, NatPhys, 2016) rather than at the X-line as expected for laminar reconnection. This case shows that the IDR can be turbulent (nonlinear), Figure 11g, and other spacecraft observations support this, e.g. observations of coalescing flux ropes (Wang+, NatPhys, 2016) and magnetic holes (Zhong+, GRL, 2019). For such nonlinear structures, available four-point measurements are not enough to reveal their topologies and the associated particle energization mechanisms.

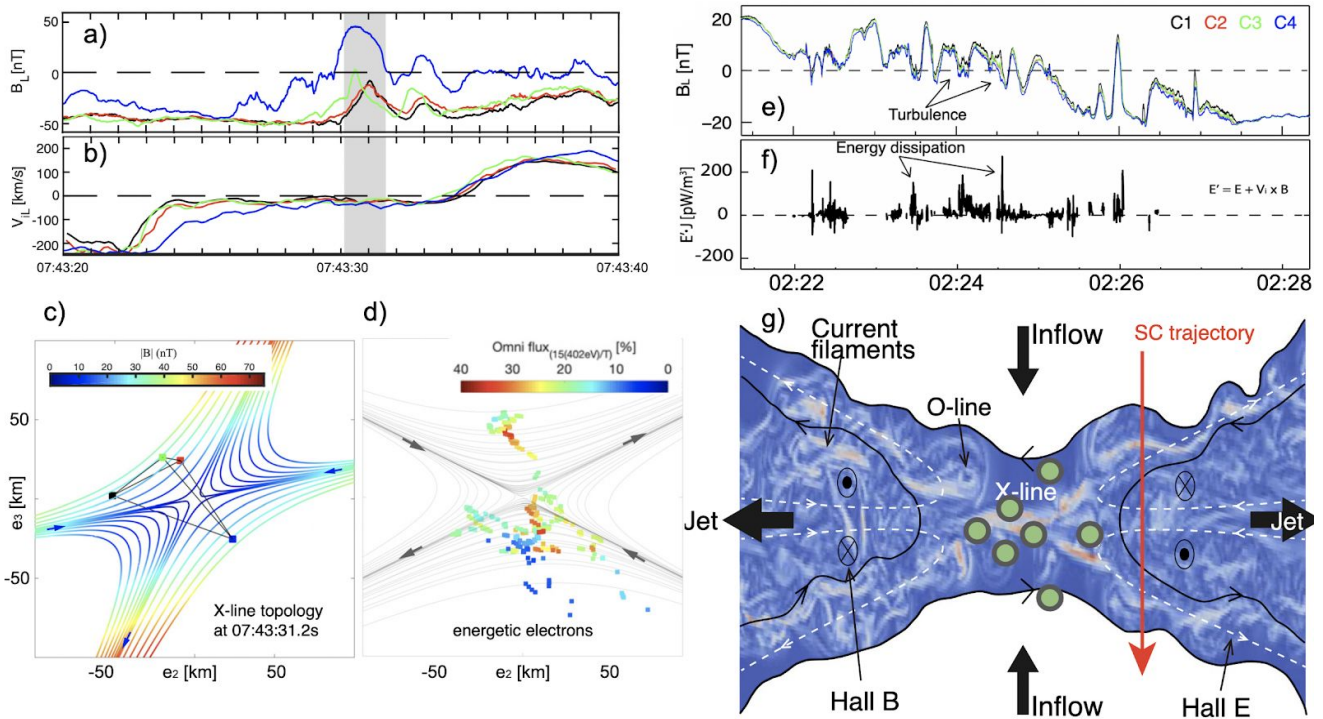

*Figure 11. Laminar EDR (Fu+, 2019) : MMS measurements of a) magnetic field and b) reconnection jet reversal. Results form FOTE method: c) magnetic field topology d) map of electron energization. Turbulent IDR (Fu+, GRL, 2017): Cluster measurements of e) magnetic field showing current filaments and f) energy dissipation and g) cartoon showing a schematic 7 spacecraft constellation that would allow to resolve the spatial and temporal evolution of the turbulent diffusion region and particle electron energization therein.*

At least 7 measurement points are needed to fully characterise the spatial and temporal evolution of turbulent reconnection and associated particle energization structures. Future multi-point observations in the magnetotail will be very important to help understand turbulent reconnection in other plasma environments, e.g. in the solar corona where reconnection current sheets can become turbulent forming many magnetic islands efficient for particle acceleration (Shibata+, EPS, 2001; Wan+, PoP, 2013).

### 3.5.2 Reconnection in shocks

Another science case demonstrating the need of new multi-scale measurements is particle energization during reconnection in shocks. Numerical simulations of high Mach number shock show strong electron energization during reconnection in the shock transition, with electrons being accelerated when colliding with reconnection jets and magnetic islands (Matsumoto+, Science, 2015). More recent simulations show that reconnection and energization occurs for both ions and electrons and are concentrated in magnetic islands (Bessho+, GRL, 2019). Very recent MMS case study observations have confirmed that reconnection can occur in the shock transition layer. In one case, reconnection is observed in an ion-scale current sheet having an ion jet and the typical signatures of electron-ion coupling of Hall reconnection (Wang+, GRL, 2019). In the second case, on the other hand, reconnection is observed in an electron-scale current sheet having an electron jet only but no coupling to ions (Gingell+, GRL, 2019). Electron energization is observed in both cases while ion energization is not. These results suggest that a complex coupling between ion and electron scales exist during reconnection at shocks, which is not understood. MMS measurements are able to resolve energization at electron scales however they cannot provide simultaneous measurements at ion scales. Cluster and THEMIS measurements lack the resolution to resolve electron physics. At least 7 measurement points are necessary to capture the coupling between electron and ion scales during reconnection at shocks.

These new multi-scale measurements would shed light on the role of reconnection for particle energization at shocks and its relative importance compared to other classical shock-related mechanisms such as energization due to a potential drop and wave-particle interactions. This may have a very important impact on understanding energization in solar and astrophysical shocks.

## 3.6 Added value observatory science

It is very important that part of the observational time of the plasma observatory is dedicated to other topics, e.g through guest investigator programs. This would serve the much wider science community. Earlier missions show that many discoveries are incidental and are in areas for which the missions were not originally designed. We give a few examples of important topics that may be addressed outside main science questions.

Measurements by the plasma observatory combined with ground measurements would have a major impact on understanding the magnetosphere-ionosphere coupling, such as magnetopause and jet braking region coupling to the ionosphere, auroral arc generation interaction and others. In the ionosphere global scale processes can be covered by [SuperDARN](#) radars, ground-based magnetometer networks and global imaging ([SMILE](#) mission) while high resolution multi-point localized measurements can be obtained using auroral cameras (e.g. ALIS 4D), and high resolution 3D incoherent radar observations (e.g. next generation European [EISCAT 3D](#)). This would allow for the first time to combine in situ measurements at different scales simultaneously with corresponding multi-scale ground based observations leading to many discoveries. The plasma observatory will also have an impact on understanding Ionosphere-Thermosphere coupling, since particles accelerated in the magnetosphere precipitate along the field-lines and provide important input to the ionosphere-upper atmosphere system.

Perturbed conditions in space environment, e.g. energetic particles, can have harmful impacts on essential technological systems and human health. The importance of Space Weather is recognized worldwide and large investments are made to advance the forecasting capabilities and to advance the understanding of fundamental particle energization mechanisms. The plasma observatory would allow to address many space weather science topics, e.g. the generation of Geomagnetically-Induced Currents (GICs) that can damage infrastructure on the ground and the production of very energetic particles that can damage infrastructure in space.

There are examples of high-risk, high-gain topics that the plasma observatory could also enable. For example, Cluster and MMS missions have used wave telescope methods that in the case of 4-spacecraft can distinguish the wave modes and their propagation directions (Paschmann+, 2008). With a 7-spacecraft constellation one could go a big step further using the additional spacecraft to triangulate the source regions. This would open a totally new experimental base for studying wave-particle interactions and particle energization sites. Another example could be using multi-spacecraft measurements of naturally occurring plasma waves for tomographic reconstructions of plasma density maps within the volume of constellation, similar to passive tomography methods as used in other fields, e.g. passive seismic tomography.

## 3.7 Broader Impact

There are many collisionless plasma environments in the heliosphere for which understanding particle energization processes is essential but the detailed observations needed to resolve these processes, such as those available in the near-Earth space, will not be possible in the foreseeable future. Examples are induced magnetospheres at minor bodies, magnetospheres of outer planets, the interaction region between solar wind and interstellar medium and many others. Results from the plasma observatory would be essential for understanding all such plasma environments.

In most plasma environments in the Universe, on the other hand, in situ measurements are not possible at all. However, plasma conditions in dimensionless parameter space can be similar to those in near-Earth plasmas, so that results from the plasma observatory could be exported to distant plasmas. The closest example is the lower part of the solar corona, where major particle energization occurs e.g. during solar flares. More distant examples include the interstellar medium, stars and astrospheres, supernova remnants, accretion disks and astrophysical jets, galaxy clusters and the intergalactic medium. In all these environments, major particle energization occurs as witnessed by cosmic rays and radiation that we remotely detect by telescopes. The quantitative measurements from the plasma observatory would allow one to derive quantities such as energy transfer and heating rates, energy partition between electrons, protons and heavy ions, and scaling laws as a function of key dimensionless parameters such as Mach number and plasma beta. All this would allow models of distant plasma objects to be tested, e.g. perform comparative studies with exoplanet magnetospheres and with astrophysical shocks.

# 4 Possible space mission profiles

## 4.1 Science and mission requirements

*Particles.* Core to the particle acceleration are the particle distribution functions, which in a collisionless plasma can be far from Maxwellian. Thus 3D phase space distributions of thermal electrons and mass-resolved ions need to be measured at the various scales simultaneously. The cadences are set by the scale dynamics. The measurements shall capture most of the heating and the most numerous accelerated populations. Measurements of energetic electrons and ions, together with ionic species, need to be made by one, and preferably more, spacecraft to follow the acceleration processes to higher energies and to study the role of composition as both an input and output of the acceleration mechanisms.

*Fields.* Variability and fluctuating fields are thought to be the primary acceleration agents beyond the lowest order DC thermalisation. The key measurements include DC B field for resolving energization processes ordered by ambient magnetic field, DC E field for measuring large-scale particle drifts and adiabatic energization processes, fluctuating B fields shall be measured at least up to electron cyclotron frequencies and E fields up to at least the plasma frequency simultaneously at multiple scales to determine the high-frequency wave modes and intermittency of turbulence that scatters and potentially selectively accelerates particles.

*Key science regions.* They are driven by the science questions. The key science regions are the Earth's bow-shock, magnetosheath and foreshock to study the shock processes and turbulence, also combination of those processes with reconnection. Addressing such mechanisms as diffuse shock acceleration requires that regions both close and far away (several Earth radii RE) from shock are covered. To study reconnection the key regions are magnetopause and magnetotail. For jet physics it is important to cover the magnetosheath and magnetotail, particularly the region of jet braking in the tail (around 8-10 RE from Earth) shall be covered.

*Orbit.* Optimization of orbit parameters would be needed to find the right balance between the time spent in the different key science regions. A reasonable starting point is an equatorial orbit with perigee about 8RE, to include the jet braking region, and apogee around 25 RE that is an approximate distance of reconnection sites in the magnetotail and is well outside the bow shock allowing coverage of pristine solar wind and distant foreshock.

## 4.2 Mission options

Based on the experience gathered from earlier mission concept studies, in particular Cross-Scale and THOR, and on the science results from recent multi-spacecraft data, in particular the recent from MMS, an L-class mission is needed in order to satisfy the science requirements. The mission options provided below build on experience gathered by our Team from several earlier concepts studied in the last decade, namely ESA/Cross-Scale, JAXA-CSA/SCOPE, ESA/EIDOSCOPE, ESA/THOR and ESA/PROSPERO, which all had particle energization as major science goal.

The Cross-Scale concept, which performed an ESA Phase-A study, allowed the study different constellation configurations, which eventually resulted in the choice of a 7-spacecraft constellation of identical spacecraft covering two spatial scales simultaneously by using two tetrahedrons sharing one corner. The payload was optimized in order to obtain electron resolution measurements at electron scales and ion resolution measurements at ion and fluid scales. The SCOPE JAXA-CSA concept, which had strong synergies with Cross-Scale, introduced a different approach with a constellation composed of a mother-daughter spacecraft couple carrying high-resolution payload (addressing ion and electron scales) and by 3 identical spacecraft with lower resolution payload (addressing macroscopic scales). The EIDOSCOPE concept, submitted as an ESA M3 candidate, was an additional high-resolution spacecraft completing the SCOPE constellation. The THOR ESA concept, which performed an ESA Phase-A study, was a single spacecraft concept including the highest resolution and highest quality in situ payload ever conceived. Finally the mission concept PROSPERO, submitted as an ESA F1 candidate, was a constellation of 8 cubesats and one mother spacecraft acting as a relay, which proposed the possibility to perform future high-quality in situ measurements by using smallsats.

### 4.2.1 L-class concepts

Different L-class mission concepts can be envisaged. We present two concepts, both based on earlier heritage:

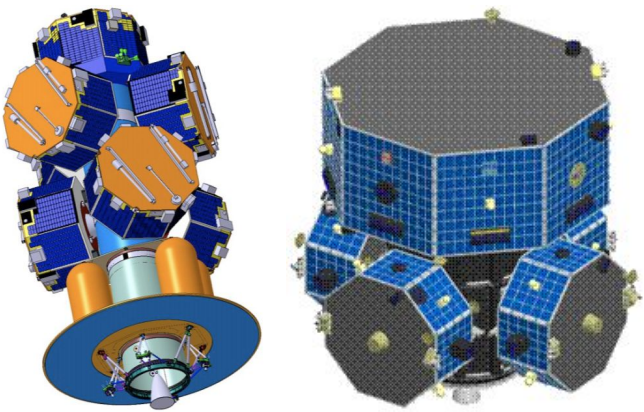

*Figure 12. Spacecraft in launch configuration including dispenser. Left: ESA Cross-Scale 7 spacecraft. Right: JAXA SCOPE mother and 4 daughters.*

**Constellation of 7 spacecraft.** This concept would be an optimization of the earlier ESA Cross-Scale concept that is discussed in the Cross-Scale Assessment Study Report. The constellation suggested here would consist of 7 spacecraft with identical platforms and possibly two different types of payload, one tailored for electron scales and the other for ion/fluid scales. The actual repartition between the two types of payload will depend on the technological and scientific developments. In the case of Cross-Scale the payload on each spacecraft was of the order 25-30kg and the spacecraft dry mass was of the order 200kg. There was a transfer module to bring the spacecraft to the final orbit. Therefore, the requirements on the dV for the spacecraft themselves were small, of the order 125m/s to keep the constellation in required formations. The wet mass of each spacecraft was only about 250kg. Altogether the wet mass of spacecraft and transfer module including margins were about 3500kg. While it was optimized for Soyuz, the Ariane launcher would allow larger launch mass and thus more payload on each of the spacecrafts. As a rough estimate we can assume at least 40kg payload on each spacecraft.

There are several parts of the Cross-Scale approach that can be further optimized, where of particular importance is the payload capacity. For example, using sun-pointing spacecraft would allow for higher accuracy electric field and particle measurements, as well as the possibility of including dedicated solar wind instruments such as a Faraday cup or a cold solar wind ion electrostatic analyzer. Another example is increasing the relative contribution of mass-resolved ion instruments due to the high importance of understanding the mass dependence of particle acceleration processes. The orbit can also be optimized, for example using lower perigee, 8 instead of 10 Earth's radii, to better cover the jet breaking and particle injection processes.

**Mother and 6 daughters**. This would be an optimization of the earlier JAXA-CAS SCOPE concept: a constellation of one mother spacecraft and 6 identical daughters all carrying identical scientific payload. The mother spacecraft will have a SCOPE-like or THOR-like high-resolution payload, while the 6 smaller daughters will have lower resolution payloads compared to the mother spacecraft. Such a setup would allow some of the instruments that do not require 7-point measurements to be put only on mother, while the daughters could include instruments for key particle and field measurements. Based on heritage and assuming an Ariane launcher we can make a rough estimate of 150kg payload on the mother and 30kg payload on each of the daughters. Depending on the outcome of technological studies, this concept may require mother-daughter communication such that all ground communication would go via the mother spacecraft. This would simplify ground segment but tradeoff studies are needed to see if it outweighs the complications introduced by the inter-spacecraft communication.

**Payload.** To address science requirements currently we expect that all 7 spacecraft should have electric and magnetic field (fluxgate and search-coil) instruments, electron and mass-resolved ion electrostatic analyzers and energetic particle instrument. For optimal science performance it is required that all 7 spacecraft would be spinners. The phase A study would require detailed study of tradeoffs among such things as spacecraft spin rate, number of instrument units, length of booms, size of sensors etc. Recent missions, such as MMS, THEMIS, Van Allen Probes, and phase A studies of Cross-Scale and THOR give confidence that optimal solutions can be found. As an example, the Phase-A study of the THOR mission addressed the limitations of the MMS payload and proposes several improvements for measurements at kinetic ion and electron scales, which are those requiring the highest resolution. Results are included in the THOR Assessment Study Report. The mission payload requires that important attention is paid to the EMC questions in designing a mission. However, they can be properly addressed, as demonstrated by the THOR EMC group, and past missions.

**Number of spacecraft.** While 7 points of measurements is the minimum required to get closure on the science questions presented in this White Paper, the possibility to have more points of measurements would allow to increase the science return of the plasma observatory. As an example, having 10 points of measurement would allow the study linear/stationary of structures at electron, ion and fluid scales simultaneously (by using a corner-sharing tetrahedra option) or to obtain for the first time direct estimates of the second order terms in the Taylor expansion (SOTE) method (Liu+, 2019). As another example, a constellation of 12 spacecraft would allow for 3 nested tetrahedra coveringelectron, ion and fluid scales simultaneously without the need to share one corner, which is an optimal constellation to address the three-scale coupling.

### 4.2.2 Possible M-class concept

There are possible ways to address parts of the science questions by using M-class missions. One possible example is an upgraded version of the PROSPERO concept, that was submitted as ESA F1 mission but deemed too expensive and thus fitting better the M-class envelope. F-class PROSPERO was a constellation of one mother spacecraft acting as a relay and 8 identical daughters (12U cubesats) carrying the scientific payload. In contrast to the F-class concept, the M-class version would include some instrumentation on the mother, (e.g. the ion-mass spectrometer and the cold solar wind instrument) and the daughters would be larger allowing higher resolution measurements than in the F-class version. This upgraded concept has been accepted in 2019 for a Phase 0 study at CNES (Leader. A.Retinò). This M-class option would allow the science questions of the L-class mission to be partially answered, although it would be more a dedicated plasma experiment rather than a plasma observatory serving the broad space plasma community. The availability of 8 points of measurements of electric and magnetic fields, which with current technology provide already high resolution, would establish the electromagnetic structure of particle energization regions beyond the linear and steady approximations, which will be an observational first. By using these measurements, important questions related to the coupling between scales would also be answered. On the other hand, those questions requiring high resolution particle measurements, e.g. the identification of energization mechanisms and the quantification of energy partition among energy range and species, would be only partially answered due to the current limitations of particle detectors, which require high mass and power in order to address kinetic scales (in particular electrons). Yet, success in some of the technological developments discussed above, would reduce these limitations and could allow to answer a significant part of the L-class science.

### 4.2.3 Possible technological developments and impact on mission options

The most important areas of technological development include *at the constellation level*: optimization of industrial production and AIT/AIV activities of many spacecraft, Inter Spacecraft Link and ranging, automatisation of mission and science operations, *at the spacecraft level*: platform design allowing flexible payload options, utilisation of smallsats, *at the payload level*: miniaturisation of plasma instrumentation and simplification of payload operations.

**Smallsats.** The recent COSPAR Roadmap on [Small Satellites for Space Science](#) has highlighted the importance of smallsats for scientific research in upcoming decades. Obtaining high scientific quality measurements using smallsats require miniaturization efforts of different subsystems, e.g. AOCS, including payload miniaturization while keeping its performance. A careful study of EMC on small platforms is needed since the physics of particle energization requires good quality of electromagnetic fields and distribution functions measurements.

**Inter Spacecraft Link (ISL).** For the concepts where daughter spacecraft communicate with the Earth through mother, it is important that ISL has sufficiently high bit rate for expected spacecraft distances.

**Ranging**. Accuracy of ranging should be carefully addressed since this has an impact on the measurement errors.

**Payload miniaturization.** The goal is to reduce the mass, volume and power consumption, which is important in general when integrating many instruments on a platform and is of particular importance for smallsat platforms. We cite here several examples. Miniaturized DC magnetometers have already been tested in flight onboard a cubesat in LEO and could measure the absolute magnetic field within 1 % (Archer+, AnGeo, 2015) and developments are continuing. Short and low-weight high-frequency search-coil magnetometers are being developed e.g. at LPP by optimizing the shape of the ferromagnetic core (Coillot+, SensLett, 2007). Developments are ongoing e.g. at IRAP, France to make new compact instruments that measure thermal ions and electrons with the same head, from a few eV to 35 keV. A new generation of electrostatic analyzer with a "donut shape" has been designed at LPP for providing an instantaneous field of view of $2\pi$ steradians and for measuring distribution functions with a very high time resolution (Morel+, JGR, 2017). Such a design is an alternative to earlier MMS and other spacecraft strategies and would lead to major savings in mass and power consumption. A miniaturized detector for electrons in the energy range 50 keV to 500 keV and protons from about 100 keV to over 6 MeV based on a solid-state telescope structure, allowing for particle species identification and energy measurement is also being developed for a CubeSat mission in the Finnish Centre of Excellence in Sustainable Space. From the detection electronics point of view, many developments are ongoing to miniaturize front-end electronics in the form of ASICs, e.g. for electron and ion detectors at LPP and IRAP. Similarly, ASICs have been developed at LPP for the front-end electronics of search coil magnetometers, such as the one onboard the JUICE mission.

**Payload industrial production.** Due to the large number of identical instruments to be produced, assembled and tested, new strategies are currently being discussed in order to reduce the cost and complexity while keeping the same quality. New strategies for science calibration of many identical instruments are also being defined.

**Spacecraft and science operations.** Technological developments are needed to make the spacecraft and science operations more autonomous and therefore less expensive. For example, more flexible schemes of controlling the quality of the spacecraft constellation could be developed. Developments are expected to reduce the complexity of science operations by using onboard automatic algorithms e.g automatic selective downlink.

Many R&D activities related to the technological developments discussed above are ongoing or are planned for the upcoming years in Europe and involve industry, national space agencies and laboratories in many universities and research centers. Overall, all this could lead to a substantial reduction in complexity and costs making such a 7-spacecraft plasma observatory possible.

## 4.3 Ground segment support

### 4.3.1 Mission and science operations

Despite the large number of spacecraft and instruments, mission and science operations of the plasma observatory could be kept relatively simple when compared to e.g. Cluster, Solar Orbiter or JUICE, as for example demonstrated by the MMS mission (Fuselier+, SSR, 2016). Missions with many instruments having very different scientific objects require very complex planning of spacecraft and payload operations. In the case of the plasma observatory, mission operations can be kept simple because the payload operates as a single instrument. There are further ways to simplify operations: spacecraft operate in a limited number of modes, e.g science, calibration, and stand-by modes, autonomous handling of the scientific data acquisition including automatic selective downlink that is currently operated manually e.g. on MMS.

### 4.3.2 Scientific data analysis

The wealth of existing 4-spacecraft analysis from Cluster and MMS, including timing analysis of plasma boundaries and other structures and magnetic field and plasma spatial gradients, have clearly shown the limitations of existing observations, particularly the strict assumption on linear gradients or 1D-structure assumption that is implicit in most 4-spacecraft methods.

With 7 spacecraft, most of these limitations can be overcome and both nonlinear analysis (e.g. determination of non-linear gradients) and identification of temporal evolution between measurement points can be addressed. Beyond the number of spacecraft, particular attention should be paid to the accuracy of measurements as non-linear analysis methods are more sensitive to measurement errors.

A number of extensions to more than 4 measurement points already exist (e.g. Paschmann+, 1998; Paschmann+, 2008; Blagau+, AnnGeo, 2010) and others are ongoing e.g. to improve the reliability of the estimated gradients and wave vectors for constellations with any number of spacecraft (Chanteur+, 2016). The space plasma community has gathered in the last two decades much experience with 4-spacecraft methods from Cluster and MMS, which makes the analysis of 7-point measurements from several instruments, although complex, completely feasible.

### 4.3.3 Numerical simulation support

The boost in the last two decades of supercomputer simulations have made the theoretical and experimental space plasmas communities strongly interconnected. This has resulted in dedicated numerical simulation support teams in recent space plasma missions, such as for MMS (Hesse+, SSR, 2016). Beyond providing theoretical predictions on particle energization, simulations provide valuable support to mission design and payload. For the 7-spacecraft observatory, one can make a virtual constellation "flying" through the simulation box in order to find the optimal location and the inter-spacecraft separation for each of the science questions.

An example in Figure 13 shows a Cross-Scale like constellation "flying" through a 3D kinetic supercomputer simulation of reconnection. Reconnection starts uniformly but quickly instabilities arise and cause the initiation of turbulence both in the inflow region near the separatrices and in the outflow jets. The ion and electron-scale tetrahedra capture very different structure in the current, both in the central diffusion region and in the two outflows. The two outflows are especially important and reveal electron scale current peaks that have an internal structure even at the electron scale. These are regions where the turbulence has reached the electron scale, producing electron-scale currents that may dissipate energy via electron-scale reconnection.

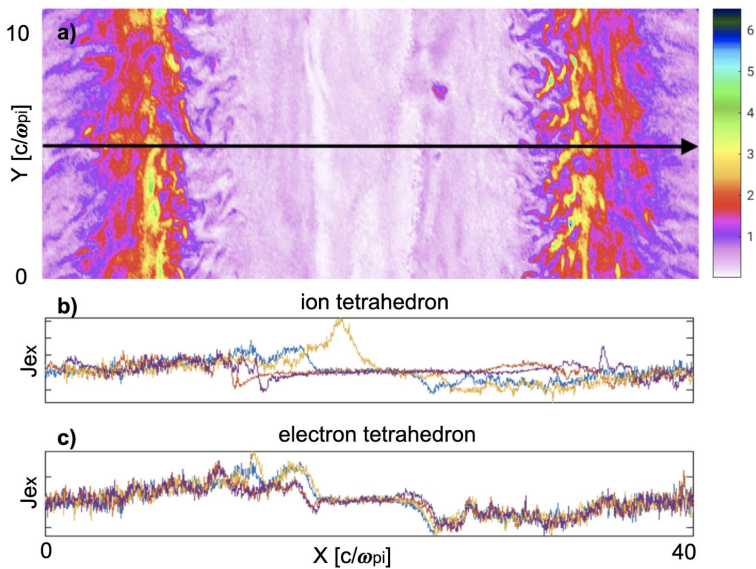

*Figure 13. Cross-Scale like constellation of seven virtual spacecraft "flying" through a supercomputer simulation of reconnection: a) cross section of the simulation along the current sheet showing the electron current in code units. Black line shows the trajectory of the virtual spacecraft. Measurements of currents by 7 spacecraft b) from the ion-scale tetrahedron and c) from the electron-scale tetrahedron (one spacecraft is common). This simulation is a development of the one in Lapenta+, NatPhys, 2015 but uses more resolution.*

## 5 Worldwide context

International collaborations are crucial for implementing the plasma observatory enhancing its capabilities and scientific return by providing additional resources and expertise. Examples of past collaborations exist both at spacecraft and payload level, such as large programmatic missions e.g. Cluster, which was together with SOHO one of the cornerstone missions of the ESA Horizon 2000 program, and MMS, as well as the Cross-Scale and SCOPE Phase-A mission studies.

Important synergies can be envisaged using coordinated measurements between the plasma observatory, other missions covering other regions of the near-Earth space and future ground observatories. This would lead to a coordinated and simultaneous coverage of near-Earth space at scales from the tiny electron scales to the global magnetospheric ones. This has never been done before and would have major scientific impacts, e.g. on space weather. Many discussions on such synergies are ongoing between scientists and we encourage more such discussion also at the level of space agencies.

In France, the need for a multi-point, multi-scale constellation addressing fundamental plasma processes has been identified as a top priority by the CNES Advisory Group on Sun, Heliosphere and Magnetospheres. A mission concept consisting of a constellation of smallsats in near-Earth space to study particle energization, based on the PROSPERO concept, has been selected by CNES for a Phase-0 study.

Italy has a large community in laboratory plasmas, space plasmas and space weather for which the science questions of particle energization is of high importance. Italy had important involvement in the earlier multiscale mission concepts Cross-Scale, EIDOSCOPE and PROSPERO. Moreover, the ASI's "Italy's Roadmap towards Space Weather science" recognizes the importance of in situ plasma multipoint observations.

In the United States, approximately 25 community white papers are associated with a constellation mission concept. As an example, the mission concept [HELIOSWARM](HELIOSWARM) will be proposed to the 2019 MIDEX call. It is a swarm of one mother spacecraft and 8 smallsats, similar to the PROSPERO concept. HELIOSWARM is tailored to study turbulent fluctuations in the solar wind, which are a major mechanism of particle energization, and will explore multiple spatial scales from fluid to sub-ion. The HELIOSWARM mission includes a strong European participation both in terms of science and instrumentation. Another 12-spacecraft constellation concept called MagCon is tailored to study the coupling between the magnetosphere and the ionosphere, particularly in the region where plasma jets interact with the inner magnetosphere. MagCon would measure at much larger spatial scales (~several Earth's radii) and the synergy between the plasma observatory and MagCon would allow the unique opportunity to simultaneously cover kinetic, fluid and global scales.

In Japan, a SCOPE Working Group was active 2003-2014. Despite the SCOPE mission not being finally realized, SCOPE science questions, which are very close to those in this White Paper, are still of major priority for the Japanase community. In order to address this science using Japanese small launch vehicle Epsilon, a mission called NEO-SCOPE is under consideration. NEO-SCOPE consists of one 200kg-size mother spacecraft and at least four 50kg-size micro spacecraft that can make full plasma measurements. As an option, adding micro-spacecraft with an X-ray imager is also considered in order to realize simultaneous measurement of global structure of the magnetosphere. NEO-SCOPE is also a testbed for the future formation flying micro-satellite missions to planetary environments.

In China, since 2016 the National Space Science Center (NSSC) has started to carry out concept studies on a mission proposal named "Self-Adaptive Magnetic Reconnection Explorer" (SAME). SAME targets the cross-scale science of magnetic reconnection with a fleet of 12+ cubesats and one mother satellite. The SAME proposal is in the pre-phase-A stage in the Strategic Priority Research Program on Space Science II, Chinese Academy of Sciences (CAS). This program has supported a series of space science missions, such as the joint ESA-CAS SMILE mission (2023) to explore the solar-wind-magnetosphere interactions. The solar-terrestrial connection is one of the preferred themes supported by the Strategic Priority Research Program.

In Russia, the topic of particle energization is of the main national science priorities for the future space plasma program. The mission Spectr-R (2011-2019) was designed to study very fast variations below the ion time scale in the solar wind and has been highly successful. The upcoming mission Resonance-MKA is a single spacecraft mission in preparation for another upcoming mission "Resonance", to be launched 2025 or later. It is a 4-spacecraft constellation to study wave-particle interactions in the outer radiation belt. Possible future options of close spacecraft constellations (7 spacecraft or more) for multiscale studies or magnetospheric constellations of several tens of spacecraft for studies of global dynamics are considered.

The Canadian space plasmas community had a major role in the SCOPE project for which Canada was supposed to provide the three spacecraft of the outer-formation. Canada has also had an important participation in earlier multiscale mission concepts Cross-Scale and EIDOSCOPE. The Canadian community is a strong leader in ground-based observations (GeoSpace Monitoring network, 3D phased array radars, etc.) and important synergies can be expected combining those with in situ observations.

# 6 Bibliography

Alexandrova, O., et al, 2013. Solar Wind Turbulence and the Role ... Space Sci. Rev. 178, 101–139.
Angelopoulos, V., 2008. The THEMIS Mission. Space Sci. Rev. 141, 5–34.
Angelopoulos, V., et al., 2013. Electromagnetic Energy Conversion at ... Science 341, 1478–1482.
Archer, M.O., et al., 2015. The MAGIC of CINEMA: first in-flight ... Ann. Geophys. 33, 725–735.
Armstrong, J.W., et al., 1995. Electron density power spectrum in the local ... Astrophys. J. 443, 209.
Arzoumanian, D., et al., Characterizing interstellar filaments with *Herschel* ... Astron. Astrophys. 529, L6.
Aschwanden, M.J., 2005. Physics of the Solar Corona … Springer Berlin Heidelberg, Berlin, Heidelberg.
Ashour-Abdalla, et al., 2011. Observations and simulations of non-local .... Nat. Phys. 7, 360–365.
Aurass, et al., 2004. Radio Observation of Electron Acceleration at Solar ... Astrophys. J. 615, 526–530.
Bahcall, et al., 1995. Hubble Space Telescope and MERLIN Observations of ... Astrophys. J. Lett. 452, L91.
Balogh, A., et al., 2013. Microphysics of Cosmic Plasmas... Sci. Rev. 178, 77–80.
Baumjohann, W., et al., 1990. Characteristics of high-speed ion flows in ... J. Geophys. Res. 95, 3801.
Behlke, R., et al., 2003. Multi-point electric field measurements of ... Geophys. Res. Lett. 30, 1177–1.
Bell, A.R., 1978. The acceleration of cosmic rays in shock ... – I. Mon. Not. R. Astron. Soc. 182, 147–156.
Berné, O., et al., 2010. Waves on the surface of the Orion molecular cloud. Nature 466, 947–949.
Bessho, N. et al., 2019. Magnetic reconnection in a quasi-parallel shock:... Geophys. Res. Lett. 0.
Blagau, A., et al., 2010. A new technique for determining orientation and ... Ann. Geophys. 28, 753–778.
Brandenburg, A., et al., 2011. Astrophysical turbulence modeling. Rep. Prog. Phys. 74, 046901.
Bruno, R., et al., 2013. The Solar Wind as a Turbulence Laboratory. Living Rev. Sol. Phys. 10, 2.
Burch, J.L., et al., 2016. Magnetospheric Multiscale Overview and Science ... Space Sci. Rev. 199, 5–21.
Burch, J.L., et al., 2016. Electron-scale measurements of magnetic reconnection ... Science 352, aaf2939.
Caprioli, D., et al., 2015. Simulations and Theory of Ion Injection at .... Astrophys. J. Lett. 798, L28.
Cassak, P.A., et al., 2008. From Solar and Stellar Flares to Coronal Heating .... Astrophys. J. 676, L69–L72.
Cattell, C., 2005. Cluster observations of electron holes in association with ... J. Geophys. Res. 110.
Chanteur, G.M., et al., 2016. Optimal Weighting of ... AGU Fall Meet. Abstr. 21, SM21A-2448.
Chasapis, A., et al., 2015. Thin current sheets and associated electron heating ... Astrophys. J. 804, L1.
Chasapis, A., et al.., 2017. Electron Heating at Kinetic Scales in Magnetosheath ... Astrophys. J. 836, 247.
Chasapis, A., et al., 2018. In Situ Observation of Intermittent Dissipation at ….. Astrophys. J. 856, L19.
Che, H., et al., 2011. A current filamentation mechanism for breaking magnetic ... Nature 474, 184–187.
Chen, B., et al., 2015. Particle acceleration by a solar flare termination shock. Science 350, 1238–1242.
Chen, C.H.K., 2016. Recent progress in astrophysical plasma turbulence from solar ... J. Plasma Phys. 82.
Chen, C.H.K., et al., 2019. Evidence for electron Landau damping in space plasma ... Nat. Commun. 10.
Chen, L.-J. et al., 2008. Observation of energetic electrons within magnetic islands. Nat. Phys. 4, 19–23.
Chen, L.-J., et al., 2016. Electron energization and mixing observed … Geophys. Res. Lett. 43, 6036–6043.
Chen, L.-J., et al., 2018. Electron Bulk Acceleration and Thermalization ... Phys. Rev. Lett. 120, 225101.
Chian, A.C.-L., Muñoz, P.R., 2011. Detection of current sheets... Astrophys. J. 733, L34.
Coillot, C., et al., 2007. Improvements on the … Sensor Letters, Volume 5, Number 1, pp. 167-170(4)
Collinson, G., et al., 2018. New Results From *Galileo* 's First Flyby ... Geophys. Res. Lett. 45, 3382–3392.
Cozzani, G., et al., 2019. In situ spacecraft observations of a structured ... Phys. Rev. E 99, 043204.
Daughton, W., et al., 2011. Role of electron physics in the development ... Nat. Phys. 7, 539–542.
De Keyser, J., et al., 2003. Structural analysis of periodic surface waves ... Planet. Space Sci., 51, 757–768.
Dimmock, et al., 2019. Direct evidence of nonstationary collisionless shocks in .... Sci. Adv. 5, eaau9926.
Divin, A. et al., 2015. Lower hybrid drift instability at a dipolarization ... J. Geophys. Res. 120, 1124–1132.
Drake, J.F., et al., 2009. A Magnetic Reconnection Mechanism for Ion ... Astrophys. J. 700, L16–L20.
Drake, J.F. et al., 2010. A magnetic reconnection mechanism... Astrophys. J. 709, 963–974.
Egedal, J., Daughton, W., Le, A., 2012. Large-scale electron acceleration by ... Nat. Phys. 8, 321–324.
Elmegreen, et al., 2004. Interstellar Turbulence I .... Annu. Rev. Astron. Astrophys. 42, 211–273.
Eriksson, E. et al., 2016. Strong current sheet at a magnetosheath... J. Geophys. Res. 121, 2016JA023146.
Eriksson, S. et al., 2016. Magnetospheric Multiscale observations ... Geophys. Res. Lett. 43, 5606–5615.
Escoubet, C.P., et al., 2001. The Cluster mission. Ann. Geophys. 19, 1197–1200.
Foullon, C. et al., 2011. Magnetic Kelvin-Helmholtz instability at the Sun. Astrophys. J. 729, L8.
Franci, L. et al., 2018. Solar Wind Turbulent Cascade from MHD to Sub-ion Scales... Astrophys. J. 853, 26.

Fu, H.S., et al., 2011. Fermi and betatron acceleration of suprathermal ... Geophys. Res. Lett. 38.
Fu, H.S., et al., 2013. Energetic electron acceleration by unsteady magnetic ... Nat. Phys. 9, 426–430.
Fu, H.S. et al., 2015. How to find magnetic nulls and … J. Geophys. Res. 120, 3758–3782.
Fu, H.S., et al.., 2017. Intermittent energy dissipation by turbulent ... Geophys. Res. Lett. 44, 37–43.
Fu, H.S., 2019. Microphysics of a magnetic reconnection is space: Energetic electrons, In preparation.
Fujimoto, M. et al., 2009. The SCOPE Mission... Proc. Int. Conf., AIP, Tokyo (Japan), pp. 29–35.
Fuselier, S.A., et al., 2016. Magnetospheric Multiscale Science Mission ... Space Sci. Rev. 199, 77–103.
Gingell, I., et al., 2019. Observations of Magnetic Reconnection in ... Geophys. Res. Lett. 46, 1177–1184.
Goodman, J. et al., 2008. Reconnection in Marginally Collisionless Accretion... Astrophys. J. 688, 555–558.
Graham, D.B., et al., 2016. Electron currents and heating in the ... Geophys. Res. Lett. 43, 2016GL068613.
Hajivassiliou, C.A., 1992. Distribution of plasma turbulence in our Galaxy derived ... Nature 355, 232–234.
Hasegawa, H., et al., 2004. Transport of solar wind into Earth's magnetosphere ... Nature 430, 755–758.
Hasegawa, H., et al., 2009. Kelvin-Helmholtz waves at the ... J. Geophys. Res. 114, A12207.
Hesse, M., et al., 2016. Theory and Modeling for the Magnetospheric ... Space Sci. Rev. 199, 577–630.
Hietala, H., et al., 2012. Supermagnetosonic subsolar magnetosheath jets and ... Ann. Geophys. 30, 33–48.
Holmes, J.C. et al., 2018. Negative Potential Solitary Structures... J. Geophys. Res. 123, 132–145.
Huang, S.Y., et al., 2017. A statistical study of kinetic-size ... J. Geophys. Res. 122, 8577–8588.
Huang, S.Y. et al., 2018. Observations of the Electron Jet Generated... Astrophys. J. 862, 144.
Innes, D.E. et al., 1997. Bi-directional plasma jets produced by magnetic... Nature 386, 811–813.
Ji, H., Zweibel, E., 2015. Understanding particle acceleration in astrophysical... Science 347, 944–945.
Johlander, A. et al., 2016. Ion injection at Quasi-parallel Shocks Seen... Astrophys. J. Lett. 817, L4.
Johlander, A., 2019. Conditions for ion acceleration at collisionless shocks. In preparation.
Jones, F.C., Ellison, D.C., 1991. The plasma physics of shock acceleration. Space Sci. Rev. 58, 259–346.
Karimabadi, H. et al., 2011. Flushing effect in reconnection: Effects... Planet. Space Sci. 59, 526–536.
Karimabadi, H. et al, 2013. Coherent structures, intermittent turbulence... Phys. Plasmas 20, 012303.
Karimabadi, H. et al., 2014. The link between shocks, turbulence, ... Phys. Plasmas 21, 062308.
Kasahara, S. et al., 2013. Asymmetric distribution... J. Geophys. Res. 118, 375–384.
Khotyaintsev, Yu.V., et al., 2011. Plasma Jet Braking: Energy Dissipation and .... Phys. Rev. Lett. 106.
Khotyaintsev, Yu.V. et al., 2017. Energy conversion... Geophys. Res. Lett. 44, 2016GL071909.
Kirk, J.G., Skjaraasen, O., 2003. Dissipation in Poynting‐Flux–dominated .... Astrophys. J. 591, 366–379.
Kistler, L.M., et al., 2005. Contribution of nonadiabatic ions to the cross-tail ... J. Geophys. Res. 110.
Kistler, L.M., et al., 2010. Cusp as a source for oxygen in the plasma ... J. Geophys. Res. 115.
Kumar, R., et al.., 2017. Preferential Heating and Acceleration of Heavy Ions ... Astrophys. J. 835, 295.
Lapenta, G., et al., 2011. Self-consistent seeding of the interchange instability ... Geophys. Res. Lett. 38.
Lapenta, G., et al., 2015. Secondary reconnection sites in reconnection .... Nat. Phys. 11, 690–695.
Lazarian, A., Vishniac, E.T., 1999. Reconnection in a Weakly Stochastic Field. Astrophys. J. 517, 700–718.
Lazarian, A., Opher, M., 2009. A model of acceleration of anomalous cosmic... Astrophys. J. 703, 8–21.
Le, A., et al., 2018. Drift turbulence, particle transport, and anomalous ... Phys. Plasmas 25, 062103.
Li, X. et al., 2018. Observing Kelvin–Helmholtz instability in solar blowout jet. Nature Sci. Rep. 8, 1, 1-9.
Liu, C.M. et al., 2018. Electron Jet Detected by MMS at Dipolarization... Geophys. Res. Lett. 45, 556–564.
Liu, T.Z., et al., 2019. Relativistic electrons generated at Earth's quasi-parallel ... Sci. Adv. 5, eaaw1368.
Liu, Y.Y., et al., 2019. SOTE: A nonlinear method for magnetic topology ..., submitted to ApJ.
Liu, Y.Y., et al, 2019. Parallel Electron Heating by Tangential Discontinuity in ... Astrophys. J. 877, L16.
Lobanov, A.P. et al., 2006. Dual-Frequency VSOP Imaging of the Jet... Publ. Astron. Soc. Jpn. 58, 253–259.
Lucek, E.A., et al., 2008. Cluster observations of the Earth's … J. Geophys. Res. 113.
Matsumoto, Y., et al., 2015. Stochastic electron acceleration during spontaneous .... Science 347, 974–978.
Matthaeus, W.H., et al., 1986. Turbulent magnetic reconnection. Phys. Fluids 29, 2513.
Matthaeus, W. H. 2015, Intermittency, nonlinear dynamics and dissipation… Phil.Trans.A 373: 20140154.
McKenzie, D.E., Savage, S.L., 2009. Quantitative examination of .... Astrophys. J. 697, 1569–1577.
McNamara, et al., 2005. The heating of gas in a galaxy cluster by X-ray cavities and ... Nature 433, 45–47.
Meyer, P., 1978. The cosmic-ray isotopes. Nature 272, 675.
Miceli, M. et al., 2019. Collisionless shock heating of heavy ions in SN 1987A. Nat. Astron. 3, 236–241.
Moore, T.W., et al., 2016. Cross-scale energy transport in space plasmas. Nat. Phys. 12, 1164–1169.
Morel, X., et al., 2017. Electrostatic analyzer with a 3-D ... J. Geophys. Res. 122, 3397–3410.

Muñoz, P.A., et al., 2018. Kinetic turbulence in fast three-dimensional collisionless ... Phys. Rev. E 98.
Nakamura, R., et al., 2009. Evolution of dipolarization in the ... Ann. Geophys. 27, 1743–1754.
Nakamura, T.K.M., et al., 2017. Turbulent mass transfer caused ... Nat. Commun. 8. Article number: 1582
Ofman, L., et al., 2011. SDO /AIA observation of Kelvin–Helmholtz instability ... Astrophys. J. 734, L11.
Oka, M., et al., 2008. Magnetic Reconnection by a Self-Retreating X Line. Phys. Rev. Lett. 101.
Pan, D., et al., 2018. Rippled Electron-Scale Structure of a ... Geophys. Res. Lett. 45, 12,116-12,124.
Panov, E.V., et al., 2013. Transient electron precipitation ... J. Geophys. Res. 118, 3065–3076.
Paschmann, G., et al., 1998. Analysis Methods for ... ISSI Scientific Reports Series SR-001, ESA/ISSI, Vol. 1.
Paschmann, G., Daly, P.W., 2008. Multi-Spacecraft Analysis Methods Revisited. ISSI Scient. Rep. SR-008
Perri, S., et al., 2012. Detection of Small-Scale Structures in the Dissipation ... Phys. Rev. Lett. 109.
Perrone, D., et al., P., 2013. Vlasov simulations of multi-ion plasma turbulence in … Astrophys. J. 762, 99.
Pezzi, O., et al., 2017. Turbulence generation during the head-on collision ... Phys. Rev. E 96, 023201.
Phan, T.D., et al., 2000. Extended magnetic reconnection at the Earth's .... Nature 404, 848–850.
Phan, T.D., et al., 2016, MMS observations of electron-scale... Geophys. Res. Lett. 43, 12, 6060-6069.
Phan, T.D., et al., 2018. Electron magnetic reconnection without ion coupling .... Nature 557, 202–206.
Plaschke, F., et al., 2013. Anti-sunward high-speed jets in the subsolar ... Ann. Geophys. 31, 1877–1889.
Plaschke, F., et al., 2017. Magnetosheath High-Speed ... J. Geophys. Res. 122, 10,157-10,175.
Plaschke, F., et al., 2018. Plasma flow patterns in and around ... Ann. Geophys. 36, 695–703.
Plaschke, F., et al., 2018. Jets Downstream of Collisionless Shocks. Space Sci. Rev. 214.
Price, L., et al., 2016. The effects of turbulence on ... Geophys. Res. Lett. 43, 6020–6027.
Pritchett, P.L., 2008. Energetic electron acceleration during multi-island ... Phys. Plasmas 15, 102105.
Pritchett, P.L., 2016. Three-dimensional structure and ... J. Geophys. Res. 121, 214–226.
Pucci, F., et al., 2017. Properties of Turbulence in the Reconnection Exhaust .... Astrophys. J. 841, 60.
Pudritz, R.E. et al., 2012. Magnetic Fields in Astrophysical Jets: From Launch... Space Sci. Rev. 169, 27–72.
Retinò, A. et al, 2007. In situ evidence of magnetic reconnection in turbulent...Nat. Phys. 3, 236–238.
Retinò, A. et al., 2008. Cluster observations of energetic electrons...  J. Geophys. Res. 113, A12215.
Retinò, A., 2016. Space plasmas: A journey through scales. Nat. Phys. 12, 1092–1093.
Reynoso, E.M. et al., 2013. On the radio polarization signature... Astron. J. 145, 104.
Rossi, C. et al., 2015. Two-fluid numerical simulations of turbulence... Phys. Plasmas 22, 122303.
Runov, A. et al., 2011. A THEMIS multicase study of dipolarization... J. Geophys. Res. 116.
Salem, C.S. et al., 2012. Identification of kinetic Alfvén turbulence... Astrophys. J. 745, L9.
Schekochihin, A.A., Cowley, S.C., 2006. Turbulence, magnetic fields,... Phys. Plasmas 13, 056501.
Schekochihin, A.A. et al., 2016. Phase mixing versus nonlinear advection in drift... J. Plasma Phys. 82.
Schwartz, S.J. et al., 1988. Electron heating and the potential jump...  J. Geophys. Res. 93, 12923–12931.
Schwartz, S.J., Burgess, D., 1991. Quasi-parallel shocks: A patchwork... Geophys. Res. Lett. 18, 373–376.
Schwartz, S.J. et al., 2009. Cross-scale: multi-scale coupling in space ... Exp. Astron. 23, 1001–1015.
Schwartz, S.J. et al., 2011. Electron Temperature Gradient Scale at Collisionless...  Phys. Rev. Lett. 107.
Schwartz, S.J. et al., 2018. Ion Kinetics in a Hot Flow Anomaly... Geophys. Res. Lett. 45, 11,520-11,529.
Sergeev, V. et al., 2009. Kinetic structure of the sharp injection/dipolarization... Geophys. Res. Lett. 36.
Servidio, S. et al., 2017. Magnetospheric Multiscale Observation of Plasma... Phys. Rev. Lett. 119, 205101.
Shay, M.A., Swisdak, M., 2004. Three-Species Collisionless Reconnection: Effect of O+... Phys. Rev. Lett. 93.
Shibata, K. et al., 2007. Chromospheric Anemone Jets as Evidence... Science 318, 1591–1594.
Shibata, K., Tanuma, S., 2001. Plasmoid-induced-reconnection...  Earth Planets Space 53, 473–482.
Shiokawa, K. et al., 1997. Braking of high-speed flows... Geophys. Res. Lett. 24, 1179–1182.
Sitnov, M. et al., 2019. Explosive Magnetotail Activity. Space Sci. Rev. 215.
Sitnov, M.I. et al., 2018. Kinetic Dissipation Around... Geophys. Res. Lett. 45, 4639–4647.
Somov, B.V., Kosugi, T., 1997. Collisionless Reconnection and High-Energy...  Astrophys. J. 485, 859–868.
Sorriso-Valvo, L. et al., 2019. Turbulence-Driven Ion Beams in the Magnetospheric... Phys. Rev. Lett. 122.
Stawarz, J.E. et al., 2019. Properties of the Turbulence... Astrophys. J. 877, L37.
Stawarz, J.E. et al., 2016. Observations of turbulence... J. Geophys. Res. 121, 11,021-11,034.
Steinvall, K. et al., 2019. Multispacecraft Analysis of Electron Holes. Geophys. Res. Lett. 46, 55–63.
Sundkvist, D. et al., 2005. In situ multi-satellite detection of coherent vortices... Nature 436, 825–828.
Sundkvist, D. et al., 2007. Dissipation in Turbulent Plasma due to Reconnection... Phys. Rev. Lett. 99.
Swisdak, M. et al., 2018. Localized and Intense Energy Conversion... Geophys. Res. Lett. 45, 5260–5267.

Taylor, M.G.G.T. et al., 2012. Spatial distribution of rolled up Kelvin-... Ann. Geophys. 30, 1025–1035.
Tong, Y. et al., 2018. Simultaneous Multispacecraft Probing... Geophys. Res. Lett. 45, 11,513-11,519.
Torbert, R.B. et al., 2018. Electron-scale dynamics of the diffusion region... Science 362, 1391–1395.
Treumann, R.A., 2009. Fundamentals of collisionless shocks... Astron. Astrophys. Rev. 17, 409–535.
Ukhorskiy, A.Y. et al., 2018. Ion Trapping and Acceleration... J. Geophys. Res. 123, 5580–5589.
Uritsky, V.M. et al., 2011. Kinetic-scale magnetic turbulence... J. Geophys. Res. 116, 9236.
Vaivads, A. et al., 2004. Structure of the Magnetic Reconnection Diffusion... Phys. Rev. Lett. 93, 105001.
Vaivads, A. et al., 2011. Suprathermal electron acceleration during... Ann. Geophys. 29, 1917–1925.
Vaivads, A. et al., 2012. EIDOSCOPE: particle acceleration at plasma... Exp. Astron. 33, 491–527.
Vaivads, A. et al., 2016. Turbulence Heating ObserveR – satellite mission proposal. J. Plasma Phys. 82,5.
Valentini, F. et al., 2016. Differential kinetic dynamics and heating of ions... New J. Phys. 18, 125001.
von Papen, M. et al., 2014. Turbulent magnetic field... J. Geophys. Res. Space Phys. 119, 2797–2818.
Walker, S.A. et al., 2017. Is there a giant Kelvin–Helmholtz... Mon. Not. R. Astron. Soc. 468, 2506–2516.
Wan, M. et al, 2013, Generation of X-points and secondary islands...Phys. Plasmas, 20, 042307.
Wan, M. et al., 2016. Intermittency, coherent structures and dissipation... Phys. Plasmas 23, 042307.
Wang, R. et al., 2016. Coalescence of magnetic flux ropes in the ion diffusion... Nat. Phys. 12, 263–267.
Wang, S. et al., 2019. Observational Evidence of Magnetic Recon... Geophys. Res. Lett. 46, 562–570.
Viberg, H. et al., 2013. Mapping HF waves in the reconnection... Geophys. Res. Lett. 40, 1032–1037.
Wilson III, L.B. et al., 2014. Quantified energy dissipation... J. Geophys. Res. 119, 6455–6474.
Wygant, J.R. et al., 2005. Cluster observations of an intense normal component... J. Geophys. Res. 100
Xiao, C.J. et al., 2006. In situ evidence for the structure of the magnetic null... Nat. Phys. 2, 478–483.
Yamada, M., Kulsrud, R., Ji, H., 2010. Magnetic reconnection. Rev. Mod. Phys. 82, 603–664.
Zhong, Z.H. et al., 2019. Observations of a Kinetic‐scale Magnetic... Geophys. Res. Lett. 46(12), 6248-6257
Zieger, B. et al., 2011. Jet front-driven mirror modes and shocklets... Geophys. Res. Lett. 38, L22103.
Zimbardo, G., 2013. A Particle Accelerator in the Radiation Belts. Physics 6, 131.

**List of relevant ISSI international teams**:
1. Magnetic reconnection and particle energization: synergy of in situ and remote observations
2. Relativistic reconnection and collisionless shocks
3. Particle Acceleration at Plasma Jet Fronts in the Earth's Magnetosphere
4. Kinetic Turbulence and Heating in the Solar Wind
5. Ion and Electron Bulk Heating by Magnetic Reconnection
6. Small scale structure and transport during magnetopause magnetic reconnection: from Cluster to MMS
7. Jets Downstream of Collisionless Shocks
8. Particle Acceleration in Solar Flares and Terrestrial Substorms
9. MMS and Cluster Observations of Magnetic Reconnection
10. Magnetic Topology Effects on Energy Dissipation in Turbulent Plasma
11. Current Sheets, Turbulence, Structures and Particle Acceleration in the Heliosphere
12. Resolving the Microphysics of Collisionless Shock Waves
13. Study of the Physical Processes in Magnetopause and Magnetosheath Current Sheets Using a Large MMS Database